\documentstyle[11pt,epsfig]{article}

\def\laq{~\raise 0.4ex\hbox{$<$}\kern -0.8em\lower 0.62
ex\hbox{$\sim$}~}
\def\gaq{~\raise 0.4ex\hbox{$>$}\kern -0.7em\lower 0.62
ex\hbox{$\sim$}~}

\def\beq{\begin{equation}}
\def\eeq{\end{equation}}
\def\bea{\begin{eqnarray}}
\def\eea{\end{eqnarray}}
\def\bean{\begin{eqnarray*}}
\def\eean{\end{eqnarray*}}

\def\vp{\varphi}
\def \vpb {{\overline {\vp}}}
\def \vpbp {{\dot{\vpb}}}
\def \rb {\overline \rho}
\def \pb {\overline p}
\def \sb {\overline \sg}
\def \H {{\cal H}}
\def \R {{\cal R}}

\def \pa {\partial}
\def \ra {\rightarrow}

\def \ti {\widetilde}
\def \la {\lambda}
\def \ls {\lambda_{\rm s}}
\def \lp {\lambda_{\rm P}}
\def \Ms {M_{\rm s}}
\def \Mp {M_{\rm P}}

\def \Da {\Delta}
\def \da {\delta}
\def \b {\beta}
\def \a {\alpha}
\def \ap {\alpha^{\prime}}

\def \Ga {\Gamma}
\def \ga {\gamma}
\def \sg {\sigma}
\def \da {\delta}
\def \ep {\epsilon}
\def \r {\rho}

\def \noi {\noindent}

\begin{document}

\begin{titlepage}

\begin{flushright}
BA-TH/03-465\\
CERN-TH/2003-257\\
NSF-KITP-04-07\\
hep-th/0401112
\end{flushright}

\vspace{0.8cm}

\begin{center}

\huge
{Cosmological perturbations \\ across a curvature bounce}\footnote[1]{This research was supported in part by the {\sl National Science Foundation} under
Grant No. PHY99-07949.}

\vspace{0.8cm}

\large{M. Gasperini$^{a,b,c}$, M. Giovannini$^d$ and G. Veneziano$^{c,d}$}

\normalsize

\vspace{0.5cm}

{\sl $^a$Dipartimento di Fisica, Universit\`a di Bari, \\ 
Via G. Amendola 173, 70126 Bari, Italy}

\vspace{.1in}

{\sl $^b$ Istituto Nazionale di Fisica Nucleare, Sezione di Bari,\\
Via G. Amendola 173, 70126 Bari, Italy}

\vspace{.1in}

{\sl $^c$ Kavli Institute for Theoretical Physics, \\
University of California, Santa Barbara, CA 93186}
\vspace{.1in}

{\sl $^d$Theoretical Physics Division, CERN, \\
CH-1211 Geneva 23, Switzerland }

\vspace*{1cm}

\begin{abstract}
\noi
String-inspired cosmologies, whereby a non-singular curvature bounce is induced by a  general-covariant,  $T$-duality-invariant, non-local dilaton potential, are used to study numerically how inhomogeneities evolve and to compare  the outcome with analytic expressions obtained through different matching conditions across the bounce.  Good agreement is found  if continuity across the bounce is assumed to hold for $\cal{R}$, the curvature perturbation  on comoving hypersurfaces, rather than for the Bardeen potential.
\end{abstract}
\end{center}

\end{titlepage}

\newpage

\parskip 0.2cm

\section{Introduction}
\label{Sec1}
\setcounter{equation}{0}

The idea that string theory, with its fundamental length scale, could resolve the big bang singularity
by effectively limiting space-time curvature has led, during the last decade, to a variety of models \cite{1}   where the big bang represents a turning (rather than an end)-point in the history of the early Universe.
All these models share the property of exhibiting a bounce in curvature, since the pre (resp. post)-bang 
eras are characterized by growing (resp. decreasing) space-time curvature. Hereafter we shall refer to this generic class of models as pre-big bang (PBB) cosmologies.

Technically speaking, the PBB phase can replace the inflationary phase of standard slow-roll inflation
since it allows the cosmologically interesting scales we observe today to start well inside the Hubble radius, and thus to
have time to homogenize through causal microphysics. An advantage of PBB cosmologies \cite{BGV}
 is that they are less sensitive to the 
so-called trans-Planckian problem since the proper frequency of all relevant quantum fluctuation starts very much below the Planck (but not necessarily the string) scale; the opposite is true in the case of slow-roll inflation, where all relevant proper frequencies are initially much above the Planck scale, formally causing linear perturbation theory to break down.
Another difference is that, while in ordinary inflationary models string (or Planck)-scale physics is washed out
by inflation (except for the  above-mentioned possible imprint on initial conditions), the opposite is the case in PBB cosmologies since the high-curvature phase, the bounce, comes there {\it at the end} of inflation.

If this latter property of PBB cosmology is appealing --in that it opens an observational window on short-distance physics--
it also makes it mandatory to understand how the bounce itself may (or may not) affect different observables.
Unfortunately, this is not an easy problem to solve since the details of the poorly understood mechanism responsible for the curvature bounce may affect, in a crucial way, how different perturbations behave across the bounce itself.
As an example, while the original investigations \cite{BGGMV} argued in favour of a strongly tilted (blue) spectrum  of adiabatic density perturbations, more recently arguments have been given \cite{EKPsp} for much flatter (and 
possibly even scale-invariant) spectra.
In toy models where the bounce is schematized as a discontinuity across a space-like hypersurface those different
predictions can be shown \cite{DV} to correspond to different ways of implementing Israel's junction conditions.
This has led to much debate in the literature lately (see for instance \cite{6}).

An alternative approach is to use, as a theoretical laboratory, 
a model of background which is everywhere smooth and  regular, 
so that we can follow  both analytically and numerically the
evolution of the different fluctuations throughout. One can then replace the smooth evolution by a sudden transition, 
and check how different prescriptions match with the ``exact" results.
We will work, in particular, in the context of a perturbative approach based on the string effective action, where it is known that curvature singularities can be
eliminated either at large curvature/coupling 
by  a combination of  higher-derivative and loop corrections \cite{2} or, in the low-curvature regime, by the contribution of a non-local dilaton potential \cite{3,4,5}. In this paper we will concentrate on this second possibility, and we will consider regular cosmological   models obtained by using a non-local potential put in by hand. Such a potential may arise, however, 
from dilatonic loop corrections in manifolds
with compact spatial sections. The corresponding non-local potential energy provides a dominant correction to the regularization of the background when an effective dimensionally-reduced coupling grows large, and leads to bouncing solutions smoothly evolving (in the Einstein frame) from accelerated
contraction to decelerated expansion. 

This paper extends the results of a
previous letter \cite{7} by investigating the frame-independence
of the results,  by performing the above-mentioned comparison
with different matching conditions, and by including a perfect
fluid source both in the background and the perturbation equations. However, we shall limit our detailed discussion of  scalar perturbations   to a
particular example of gravi-dilaton background without matter
sources. The discussion of more general examples is postponed to further work \cite{8}. 

Our analytical and numerical computations  
suggests that the scalar spectrum associated to a non-singular background can be correctly
obtained through an (approximate) matching procedure,  {\em
provided} the matching prescriptions are such as to guarantee a smooth
crossing over the bounce of the variable ${\cal R}$, representing spatial
curvature perturbations on comoving (or, at large scale, on constant energy density) hypersurfaces, and {\em not} of the variable $\Psi$,
the gauge invariant Bardeeen potential. 
 It is possible, however, that such a result is due to the fact that the bounce is triggered by a potential $V$
that depends mostly on the value of the dilaton: as such, it may not apply when
the matching is performed through a {\em singular} bounce, as it seems to
be the case in the ekpyrotic scenario \cite{9}. 

The rest of the paper  is organized as follows. 
In Section II we introduce the low-energy effective action to be used
throughout this paper and discuss, in particular, the possible origin of the non-local dilaton potential. 
In Section III we derive, both in the string and in the Einstein frame, the (non-local) field equations that follow from the above action, and  present classes of exact solutions describing 
regular homogeneous bouncing backgrounds. The last part of Section III, containing the derivation of the exact solutions with fluid sources, is not essential for the understanding of the subsequent sections, and may be skipped by those readers who are only interested in the discussion of the matching prescriptions. 
In Section IV we introduce the full set of scalar perturbation
equations in the Einstein frame, including the perturbations of the fluid sources, both in the uniform-dilaton gauge and in full gauge-invariant form. We also give explicit relations to the String-frame perturbation variables, for easier comparison with  previous  papers and results. 
In  Section V we estimate analytically, and compute numerically, the scalar perturbation spectrum (inside and outside the horizon) for a typical example of regularized background, and we discuss the possibility of reproducing the same spectrum through an appropriate matching procedure across a sudden bounce. 
Section VI is  devoted to
some general comments and concluding remarks.

\section{Non-local dilaton potential and low-energy effective action}
\label{Sec2}
\setcounter{equation}{0}

It is well known that implementing a curvature bounce in
cosmology must involve some unconventional equation of state
for the dominant source of energy. Such a possibility can arise in principle from
higher-order corrections in $\ap$, string-loop corrections, or both. Examples of such
``exit" mechanisms have been widely discussed in the literature \cite{2}.
In this paper we wish to implement a low-curvature bounce and therefore we
will not appeal to the former kind of corrections. We shall  argue, instead, that string-loop corrections,
in the presence of space compactification, can provide an efficient and appealing regularization
mechanism. 

In a $D$-dimensional manifold, if $n \leq D-1$ spatial dimensions are compactified, at energies below the compactification scale the effective lower-dimensional dynamics is controlled indeed by the ``reduced" coupling:
\beq
g_{red}^2 = g_s^2/ V_n \; ,
\label{gred}
\eeq
where $g_s^2 = e^{\phi}$ is the string coupling of the full $D$-dimensional theory ($\phi$ being the dilaton)  and $V_n$ is the $n$-dimensonal compactification volume in string units.
In standard superstring compactifications from $D=10$ to $D=4$, the reduced coupling $g_4^2 \equiv g_s^2/ V_6 $ is
nothing but the
four-dimensional gauge coupling that controls the dynamics of the ``zero-modes" with respect to the six-dimensional internal space, i.e. of the light four-dimensional degrees of freedom. As an example, it is known that the running of $g_4$ 
is determined by $g_4$ itself through the infrared
behaviour of zero-mode  loops. It is also known that $T$-duality (i.e. large-small volume duality) is preserved by loops, provided the reduced coupling
is kept  fixed.

In a cosmological set-up where the whole of space is compact, the zero modes become the homogeneous components of the various fields, and their  self-interactions should be controlled by the $D=1$ effective coupling $g_1^2 \equiv g_s^2/ V_9$.
In analogy with the previous example we may expect  loop-corrections that preserve  $T$-duality where, this time, $e^{\phi(t)}/ V_9 \equiv e^{\vpb(t)} = g_1^2$ is kept fixed. In particular, if a 
non-trivial effective potential is induced, it should satisfy such a symmetry and must therefore be of the form $V_{eff} = V(\vpb)$.

It is known  \cite{3,4,5} that such potentials  can be very useful for inducing curvature bounces.
However, for a long time, it was not clear  how to formulate such a framework in a fully general-covariant way.
This problem was solved in \cite{7}, where the following generalization of $e^{-\vpb(t)}$ to a function of all the coordinates
$x$ was proposed ($d \equiv D-1$):
\beq
e^{-\vpb(x)} =   {1\over \ls^d} \int {d^{d+1}y}\sqrt{|g(y)|}~ e^{-\vp(y)}
\sqrt{\pa_\mu \vp (y) \pa^\mu \vp(y)}
~\delta\left(\vp (x) -   \vp (y)\right) .
\label{21}
\eeq
Such a quantity, even if non-local, transforms as a scalar function of  $x$ under general coordinate transformations,  and is automatically $O(d,d)$ invariant for a string-frame (S-frame) metric with $d$ abelian isometries. 
It should be remarked, however, that such a definition of $\vpb$ is only consistent for backgrounds such that
 $\pa_\mu \vp  \pa^\mu \vp >0$. This will be always the case for all background solutions we will consider in this paper. 

In the spirit of the previous discussion, our toy model will consist of taking the tree-level, low-energy effective action of
(say heterotic) string theory, and of adding to it a suitable effective potential depending only upon $\vpb(x)$, 
while other
possible matter fields are included into the effective action $S_m$.
Our effective action will then have the form, in $d+1$ dimensions, and in the S-frame:
\beq
S= -{1\over 2\ls^{d-1}} \int d^{d+1}x \sqrt{|g|}~
e^{-\vp}\left[R+ \left(\nabla \vp\right)^2+V(e^{-\vpb})\right]+S_m.
\label{31}
\eeq
 Here $\ls= \Ms^{-1}= (2 \pi \ap)^{1/2} $ is the string
length scale, and the non-local function  $\vpb=\vpb (\vp)$ is defined in
Eq. (\ref{21}). Our conventions are: diag $g_{\mu\nu}=(+--- \dots)$, 
$R_{\mu\nu\a}\,^\b= \pa_\mu\Ga_{\nu\a}\,^\b+ \Ga_{\mu\r}\,^\b
\Ga_{\nu\a}\,^\r - \dots$, and $R_{\nu\a}= R_{\mu\nu\a}\,^\mu$.

\section{General field equations and background solutions}
\label{Sec3}
\setcounter{equation}{0}

\subsection{Field equations in the string Frame}
\label{Sec3.1}

Starting from the action (\ref{31}) the field equations are obtained, as usual, by taking functional
derivatives of the action with respect to the metric and the dilaton. By 
adopting the standard definition of the matter
stress-energy  tensor, 
\beq
{\da S_m \over \da g^{\mu\nu}(x)}= {1\over 2} \sqrt{|g|}~ T_{\mu\nu} \, ,
\label{32}
\eeq
 we note that 
\bea
&&
{\da \over \da g^{\mu\nu}(x)} \int d^{d+1}x' \left(
\sqrt{|g|}e^{-\vp}V\right)_{x'}=
-{1\over 2} \left(\sqrt{|g|}e^{-\vp}g_{\mu\nu}V\right)_{x}
\nonumber\\
&&
+{1\over \ls^d} \int d^{d+1}x' \left(\sqrt{|g|}e^{-\vp}V'\right)_{x'}
{\da \over \da g^{\mu\nu}(x)}\int d^{d+1}y \left(
\sqrt{|g|}e^{-\vp} \sqrt{ (\pa \vp)^2}\right)_{y}
\da( \vp_{x'}-\vp_y) \nonumber \\
&&
=-{1\over 2} \left(\sqrt{|g|}e^{-\vp}g_{\mu\nu}V\right)_{x} 
\nonumber\\
&&
-{1\over 2 \ls^d}
\left(\sqrt{|g|}e^{-\vp}\ga_{\mu\nu}\sqrt{ (\pa \vp)^2}\right)_{x}
 \int d^{d+1}x' \left(\sqrt{|g|}e^{-\vp}V'\right)_{x'}
\da( \vp_{x}-\vp_{x'}),
\label{33}
\eea
where 
\beq
V' \equiv {\pa V\over \pa e^{-\vpb}}=-e^{\vpb} {\pa V\over \pa \vpb}
\label{34}
\eeq
and
\beq
\gamma_{\mu\nu} = g_{\mu\nu} - 
\frac{\partial_{\mu}\vp \partial_{\nu}\vp}
{(\partial \vp)^2}.
\label{35}
\eeq
We are using the convenient notation in which an index appended to
round brackets, $( \dots)_x$, means that all quantities inside the
brackets are functions of the appended variable. Similarly, $\vp_x \equiv
\vp(x)$. 

Collecting all terms, and multiplying by $e^\vp/\sqrt{|g|}$, we 
obtain the final equation 
\beq
G_{\mu\nu} + \nabla_{\mu} \nabla_{\nu} \vp + \frac{1}{2} g_{\mu\nu}  
\left[ (\nabla \vp)^2 - 2 \nabla^2 \vp - V\right] -  \frac{1}{2} e^{-\vp} 
\sqrt{(\partial \vp)^2} ~\ga_{\mu\nu} I_1 = \ls^{d-1} e^\vp T_{\mu\nu},
\label{36}
\eeq
where $G_{\mu\nu}$ is the usual Einstein tensor, and 
\beq
I_1(x) = {1\over \ls^d}\int d^{d + 1}y \left(\sqrt{|g|}~V'\right)_y
~\delta\left(\vp_x -   \vp_y\right).
\label{37}
\eeq

We now take the functional derivative of the action with respect to the
dilaton, and define the scalar charge-density of the matter sources,
$\sg$:
\beq
{\da S_m \over \da \vp (x)}= -{1\over 2} \int d^{d+1}x' 
\left(\sqrt{|g|}~ \sg\right)_{x'} ~\da^{d+1} (x-x').
\label{38}
\eeq
The differentiation of the non-local potential gives
\bea
&&
{\da \over \da \vp(x)}\int d^{d+1}x' \left(
\sqrt{|g|}e^{-\vp}V\right)_{x'}=
-\left(\sqrt{|g|}e^{-\vp}V\right)_{x} 
\nonumber\\
&&
-{1\over \ls^d} \left(\sqrt{|g|}e^{-\vp}V'\right)_{x}\int d^{d+1}y
\left(\sqrt{|g|}e^{-\vp}\sqrt{ (\pa \vp)^2}\right)_{y} \da'(
\vp_{y}-\vp_x) \nonumber \\ 
&&
-{1\over \ls^d} e^{-\vp} \pa_\mu \left(\sqrt{|g|}\pa^\mu \vp\over \sqrt{ (\pa
\vp)^2}\right)_x  \int d^{d+1}x' \left(\sqrt{|g|}e^{-\vp}V'\right)_{x'}
\da( \vp_{x}-\vp_{x'}),
\label{39}
\eea
where $\da'$ denotes the derivation of the delta distribution with respect to its argument. 
 The last term of the above equation can be rewritten as 
\beq
- \sqrt{|g|}e^{-2 \vp}~\nabla_\mu\left(\nabla^\mu \vp \over 
\sqrt{ (\pa \vp)^2}\right) I_1=-\sqrt{|g|}{e^{-2 \vp} \over
\sqrt{ (\pa \vp)^2}}~ (\ga_{\mu\nu} \nabla^\mu \nabla^\nu \vp )~I_1,
\label{310}
\eeq
while the term containing the $\da'$ integral can be rewritten as 
\beq
\sqrt{|g|}e^{- \vp} \left({\pa V \over \pa \vpb} + e^{-\vp} V' I_2\right),
\label{311}
\eeq
where
\beq
I_2(x) = {1\over \ls^d} \int d^{d + 1} y\left( \sqrt{|g|}~ \sqrt{ (\pa \vp)^2}\right)_y
\delta'\left(\vp_x -   \vp_y\right).
\label{312}
\eeq
Collecting all terms, and multiplying by $e^\vp/\sqrt{|g|}$, we 
finally obtain  the dilaton equation to read: 
\beq
R + 2 \nabla^2 \vp - (\nabla \vp)^2 + V - {\pa V\over \pa{\bar{\vp}}} +  e^{-\vp}  \frac{\widehat{\nabla}^2 \vp}{\sqrt{(\partial \vp)^2}}  I_1 - 
e^{-\vp}  V'   I_2 = \ls^{d-1} e^\vp \sg,
\label{313}
\eeq
where we have defined 
\beq
 \widehat{\nabla}^2 \vp = \ga_{\mu\nu} \nabla^\mu\nabla^\nu \vp.
\label{314}
\eeq

Eqs.(\ref{36}) and (\ref{313}) were already given (without presenting their explicit derivation) in \cite{7}. In the absence of matter sources they are not independent, being related by
\beq
\nabla^\nu \left(\da S\over \da g^{\mu\nu}\right)=
-{1\over 2} \nabla_\mu \vp \left(\da S\over \da \vp\right). 
\label{315}
\eeq
For later applications, it may be useful to consider also a combination of the metric and dilaton equations, by eliminating the scalar-curvature term appearing in Eq. (\ref{36}). The result is 
\bea
&
R_\mu~^\nu+\nabla_\mu\nabla^\nu \vp-{1\over 2}\da_\mu^\nu\left[{\pa V\over \pa \vpb}+e^{-\vp}V'I_2\right]
+{1\over 2}e^{-\vp}\left[\da_\mu^\nu \frac{\widehat{\nabla}^2 \vp}{\sqrt{(\partial \vp)^2}} -\ga_\mu^\nu \sqrt{(\partial \vp)^2}\right] I_1 =\nonumber \\
&
=\ls^{d-1} e^\vp \left(T_\mu~^\nu+{1\over 2} \da_\mu^\nu \sg\right).
\label{316}
\eea

Let us finally stress the difference with the more conventional case in which the potential appearing in the action is a local function of the dilaton, $V=V(\vp)$. In that case, the variation of the action provides a set of scalar-tensor equations which can be directly obtained from Eqs. (\ref{36}), (\ref{313}), simply by neglecting the terms with the $I_1$ and $I_2$ integrals, and replacing in Eq. (\ref{313}) $\pa V/\pa \vpb$ by $\pa V/\pa \vp$.

\subsection{Field equations in the Einstein Frame}
\label{Sec3.2}

In this paper we will complement the analysis of \cite{7} by studying the evolution of metric perturbations directly in the Einstein frame (E-frame), where the dilaton is minimally coupled to the metric and canonically normalized. To this purpose, it is convenient to write down the field equations starting directly from the E-frame action. This is obtained from the S-frame action (\ref{31}) through the conformal rescaling (here we will denote with a tilde the E-frame variables):
\beq
g_{\mu\nu}=  \left(\ls\over \lp\right)^2 \ti g_{\mu\nu} ~
e^{2\vp/(d-1)} \; .
\label{41}
\eeq
We have explicitly introduced the Planck length scale $\lp$ to restore the conventional canonical normalization of the gravitational variables. From the action (\ref{31}) we thus obtain, up to a total derivative,
\beq
S= {1\over 2\la_P^{d-1}} \int d^{d+1}x \sqrt{|\ti g |}~
\left[- \ti R+{1\over d-1} \left(\ti \nabla \vp\right)^2- e^{2\vp
\over d-1} 
\left(\la_s\over \la_P\right)^2 V(\vpb)\right] +S_m,
\label{42}
\eeq
where $\lp$ defines the $D=(d+1)$-dimensional gravitational coupling $G_D$ as 
\beq
\lp^{1-d} \equiv  \Mp^{d-1} = {1\over 8 \pi G_D},
\label{43}
\eeq
and where
\beq
e^{-\vpb(x)} =  \lp^{-d}   \int d^{d+1}y \left(
\sqrt{|\ti g |} e^{\vp\over d-1}\sqrt{(\ti \nabla \vp)^2}\right)_y \delta(\vp_x - \vp_y) .   
\label{44}
\eeq
By introducing the (E-frame) canonically normalized scalar field $\ti \vp$:
\beq
\ti \vp =\vp \left(M_P^{d-1}\over d-1\right)^{1/2}\; ,
\label{45}
\eeq
the action can be written in canonical form as
\beq
S=  \int d^{d+1}x \sqrt{|\ti g |}~
\left[- {\ti R\over 16 \pi G_D} +{1\over 2} \left(\ti \nabla  
\ti \vp\right)^2- \ti V \right]+S_m(\ti g, \ti \vp),
\label{46}
\eeq
where
\beq
\ti V= {\la_s^2\over 2 \la_P^{d+1}}~V(\vpb)~ e^{2\ti\vp/ \sqrt{(d-1)M_P^{d-1}}}.
\label{47}
\eeq

We shall work in units $16\pi G_D=1$, i.e. $\Mp^{d-1}=2$. In this case 
eq. (\ref{45}) reduces to
\beq
\ti \vp= c~ \vp, ~~~~~~~~~~~~~~ c= \sqrt{2\over d-1}.
\label{48}
\eeq
We shall also absorb the factor $(\la_s/\la_p)^2$ into the function $V(\vpb)$, with the
understanding that now the dilaton potential is 
rescaled from string units to Planck units. The action finally becomes
\beq 
S(\ti g, \ti \vp)=  \int d^{d+1}x \sqrt{|\ti g |}~ \left[- {\ti R} +
{1\over 2} \left(\ti \nabla  \ti \vp\right)^2- \ti V \right] +S_m(\ti g, \ti \vp),
\label{49} 
\eeq
where 
\bea
&& 
\ti V(\ti g, \ti \vp)=  e^{c\ti\vp}~V(e^{-\vpb}),\nonumber \\
&&
e^{-\vpb(\ti g, \ti \vp)} = e^{c\ti\vp/2}
\int d^{d+1}y \left(\sqrt{|\ti g |}\sqrt{(\ti \nabla \ti \vp)^2}\right)_y
 \delta(\ti \vp_x -  \ti \vp_y) . 
\label{410}
\eea

In order to obtain the field equations we will now take functional derivatives of the above action with respect to $\ti g^{\mu\nu}$ and $\ti \vp$. For the matter action $S_m$ we set 
\beq
\da S_m = {1\over 2} \int dx^{d+1} \sqrt{|\ti g |} ~\ti T_{\mu\nu} ~\da \ti g^{\mu\nu}, ~~~~~~~
\da S_m = -{1\over 2} \int dx^{d+1} \sqrt{|\ti g |} ~\ti \sg ~\da \ti \vp ,
\label{411}
\eeq
and  find that, according to the definition of $\ti g$ and $\ti \vp$, the E-frame sources terms are related to the S-frame ones by the rescaling
\bea
&&
\ti T_\mu~^\nu =  T_\mu~^\nu~ e^{{d+1\over d-1} \vp}=
T_\mu~^\nu~ e^{{c \over 2}(d+1) \ti \vp}, 
\nonumber\\
&&
\ti \sg =  \sqrt{d-1\over 2} ~\sg~ e^{{d+1\over d-1} \vp}=
{\sg \over c}~ e^{{c \over 2}(d+1) \ti \vp}.
\label{412}
\eea
By separating explicitly the non-local part $V$ of the potential $\ti V$ of Eq. (\ref{410}), and following the same procedure as in the S-frame, the variation of the full action with respect to $\ti g^{\mu\nu}$ gives  
\beq
\ti G_{\mu}~^\nu={1\over 2} \ti T_{\mu}~^\nu+{1\over 2} \left (\pa_\mu  \ti \vp
\pa^\nu \ti \vp -{1\over 2} \da_\mu^\nu ~\pa \ti \vp^2\right)+{1\over 2}  e^{c \ti \vp}
\left(\da_\mu^\nu V + \ti \ga_{\mu}~^\nu \sqrt{\pa \ti \vp^2} 
~e^{c\ti \vp/2}~ J_1 \right), 
\label{413}
\eeq 
where
\beq
\ti \ga_{\mu\nu}=\ti g_{\mu\nu} -{\pa_\mu \ti \vp \pa_\nu \ti \vp\over \pa \ti \vp^2},   
~~~~~~~~~
J_1(x)={1\over \lp^d}\int d^{d+1}y \left(\sqrt{|\ti g|} V'\right)_y \da(\ti \vp_x-\ti \vp_y),
\label{414}
\eeq
and, as before, $V'= (\pa V/\pa e^{-\vpb})$. The variation with respect to $\vp$ gives 
\beq
\ti \nabla^2 \ti \vp +{1\over 2} \ti \sg +{1\over 2} c \ti T
+ e^{c \ti \vp}\left[
cV-{c\over 2}{\pa V\over \pa \vpb}
- e^{c\ \ti \vp/2}\left( \frac{\ti \ga_{\mu\nu} \ti \nabla^\mu \ti \nabla^\nu \ti \vp}{\sqrt{(\partial \ti \vp)^2}} J_1-V'J_2\right) \right]=0,
\label{415}
\eeq
where
\beq
J_2(x)={1\over \lp^d} \int d^{d+1}y \left(\sqrt{|\ti g|} \sqrt{\pa \ti \vp^2}\right)_y \da' (\ti \vp_x-\ti \vp_y).
\label{416}
\eeq 
The presence of the trace of the stress tensor, in this equation, is due to the $\vp$-dependence of the matter action through the rescaled metric $\ti g$, leading to the variational contribution 
\beq
\left( \da S_m \over \da g^{\mu\nu}\right) {\da  g^{\mu\nu} \over \da \ti \vp}= {1\over 2} \sqrt{ |g|}~ T_{\mu\nu} {\da \over \da \ti \vp}
\left(\ti g^{\mu\nu} e^{-c \ti\vp}\right) = -{c\over 2} \sqrt{|\ti g|}~ \ti T.
\label{417}
\eeq
The combination of the two equations (\ref{413}), (\ref{415}) leads to the generalized conservation equations 
\beq
\ti \nabla_\nu \ti T_\mu~^\nu-{1\over 2}\left(\ti \sg+c \ti T\right) \ti \nabla_\mu \ti \vp
=0. 
\label{418}
\eeq

We finally recall, for comparison, that the (more conventional) equations obtained from the Einstein action (\ref{49}) in which the S-frame potential $V$ is a {\em local} function of the dilaton, $V=V(\ti \vp)$, can be simply obtained from Eqs. 
 (\ref{413}), (\ref{415})  by dropping the integrated contributions  $J_1, ~J_2$, and by replacing $-(c/2) (\pa V/\pa \vpb)$ by $\pa V/\pa \ti \vp$ in Eq. (\ref{415}).

\subsection{Homogeneous background solutions }
\label{Sec3.3}

For the cosmological applications of this paper we have to start with (and then perturb) a homogeneous, isotropic and spatially flat background. Considering first the S-frame equations we set: 
\beq
g_{\mu\nu}= {\rm diag} (1, -a^2 \da_{ij}), ~~~~~~~
T_\mu~^\nu= {\rm diag} (\rho, -p \da_{ij}), ~~~~~~~
\vp=\vp(t)
\label{317}
\eeq
(we are assuming that the matter stress tensor can be expressed in the perfect fluid form), and we obtain we obtain, from Eq. (\ref{21}),
\beq
e^{-\vpb}= e^{-\vp} a^d,
\label{318}
\eeq
where, assuming spatial sections of finite volume,  we have normalized the scale factor $a$
as to absorb in $a^d$ the S-frame spatial volume in string units. 

Let us write the associated S-frame field equations, in units $2 \ls^{d-1}=1$. 
The $(0,0)$ component of eq. (\ref{36}) gives 
\beq
\dot{\overline{\vp}}^2 - d H^2 -V=e^\vpb \rb,
\label{319}
\eeq
while the $(ij)$ component of eq. (\ref{316}) gives
\beq
\dot H -H\vpbp={1\over 2} e^\vpb\left(\pb-{1\over 2} \sb\right),
\label{320}
\eeq
and the dilaton equation (\ref{313}) gives
\beq
2 \ddot \vpb - \vpbp^2 -d H^2 +V -{\pa V\over \pa \vpb} = 
{1\over 2} e^\vpb \sb.
\label{321}
\eeq
Their combination leads to the matter conservation equation
\beq
\dot {\rb} +dH\pb= {1\over 2}\sb\left(\vpbp+dH\right),
\label{322}
\eeq
where $\rb=\r a^d$, $\pb=p a^d$, $\sb=\sg a^d$.

When written explicitly in terms of $\vp$, the above equations can be easily compared with the homogeneous cosmological equations associated with a local potential $V(\vp)$. Besides the replacement $\pa V/\pa \vpb \ra \pa V/\pa \vp$ in the equation for the dilaton field, there is a second remarkable difference appearing in the spatial equation corresponding to Eq. (\ref{320}). For a local potential $V(\vp)$ such equation contains, on its left-hand side, the additional term $(1/2)(\pa V/\pa \vp)$. As illustrated below, this difference is responsible for making our equations admit exactly integrable and regular bouncing solutions. 

In the absence of matter sources, $T_{\mu\nu}=0$, $\sg=0$, the dilaton equation (\ref{321}) becomes a consequence of Eqs. (\ref{319}), (\ref{320}) (provided $\vpbp\not=0$). The whole system of background equations can then be reduced to quadratures, namely 
\beq
H = m e^\vpb, ~~~~~~~~~
t= \int^\la d\la' \left[d+\la'^2 V(m\la')\right]^{-1/2},
\label{323}
\eeq
where $m$ is an integration constant, and we have defined $\la=m^{-1} e^{-\vpb}$. As noticed in \cite{3}, it is then possible to obtain, quite generically, non-singular solutions if the function appearing between square brackets in the above integral has a simple zero.  Furthermore, if $V\la^2 \ra 0$ at large $|\la|$ (corresponding to a potential generated beyond the two-loop level), then the solutions asymptotically approach, at large $|t|$, the usual isotropic solutions dominated by the dilaton kinetic energy (see for instance \cite{1}), characterized by $H |t| = 1/\sqrt{d}$. 

To illustrate this point, and to provide explicit analytical examples of regular backgrounds, we shall consider here the particular class of potential defined by
\beq
V(m\la)= \la^{-2}\left( \left[\a -(m\la)^{-2n}\right]^{2-1/n} - d \right),
\label{324}
\eeq
and parametrized by the dimensionless coefficients $\a$ and by the ``loop-counting" parameter $n$. For $\a >0$ and $n>0$, Eq. (\ref{323}) can then be integrated exactly, and leads  to the general  solution
\beq
H =m e^\vpb =m\left[\a \over 1+(\a mt)^{2n}   \right]^{1/2n},
\label{325}
\eeq
whose ``bell-like" shape describes a bouncing evolution of the curvature scale, i.e. a smooth transition from growing to decreasing curvature, and from accelerated to decelerated expansion. 

The asymptotic behaviour of this class of solutions is controlled by the parameter $\a$. As $|t| \ra \infty$ one finds indeed $H= \a^{(1-2n)/2n} /|t|$, so that the background (\ref{325}) evolves from an initial accelerated, growing dilaton configuration,  parametrized  for $ t \to - \infty$ as 
\beq
 a \sim (-t)^{-\a^{(1-2n)/2n}}, ~~~~~~~~~
\vp \sim -\left[d~ \a^{(1-2n)/2n} +1\right] \ln (-t),  ~~~~~~~~~
t<0,
\label{326}
\eeq 
to a final ``dual" and time-reversed configuration, 
parametrized  for $ t \to +\infty$ as 
\beq
 a \sim t^{\a^{(1-2n)/2n}}, ~~~~~~~~~
\vp \sim \left[d~ \a^{(1-2n)/2n} -1\right] \ln t,  ~~~~~~~~~
t>0. 
\label{327}
\eeq 
The minimal, pre- and post-big bang solutions dominated by the dilaton kinetic energy \cite{1} are thus recovered, asymptotically, for
\beq
\a= d^{n/(2n-1)}.
\label{328}
\eeq
If this condition is not satisfied the background is potential-dominated even asynptotically. 

For such a class of bouncing solutions the dilaton keeps growing as $t \to +\infty$, if $\a^{(1-2n)/2n}d$ $>1$. It is thus possible that a solution, which is regular in the S-frame metric corresponding to the action (\ref{31}), acquires a curvature singularity when transformed to the 
E-frame metric corresponding to the action (\ref{46}). Let us compute, indeed, the E-frame curvature scale, or the Hubble parameter
\beq
\ti H= {d \ln \ti a\over d \ti t}= \left(H-{\dot \vp \over d-1}\right)
e^{\vp/(d-1)}, ~~~~~~~~
d \ti t=dt~ e^{-\vp/(d-1)},
\label{329}
\eeq
where $\ti t$ is the cosmic-time parameter in the syncronous E-frame gauge, while the dot still denotes differentiation with respect to the 
S-frame cosmic time parameter.  
Asymptotically, at large positive times, we obtain from Eq. (\ref{327}):
\beq
\ti H \sim \left[1- \a^{(1-2n)/2n} \over d-1 \right]  \times 
t^{~{d \over d-1}\left[\a^{(1-2n)/2n}-1\right] }.
\label{330}
\eeq
The solution is thus bouncing even in the Einstein frame (irrespective of the possible growth of the dilaton), provided 
\beq
 \a^{(1-2n)/2n} <1.
\label{331}
\eeq
This conditions guarantees that the  Hubble factor of the E-frame metric approaches zero from positive values as $t \to +\infty$, and we note, in particular, that the condition is always satisfied by the free vacuum solutions defined by Eq. (\ref{328}). 

It is important to stress, for subsequent applications to metric perturbations, that the asymptotic form of the Einstein scale factor $\ti a$, when written in conformal time, becomes completely independent of the parameters $\a$ and $n$:
\bea
&&
\ti a \sim  a e^{-\vp/(d-1)} \sim (-t)^{1+  \a^{(1-2n)/2n}   \over d-1}
\sim (-\eta)^{1\over d-1}, ~~~~~~~~ \eta<0, \nonumber\\
&&
\ti a  \sim a e^{-\vp/(d-1)} \sim t^{1-  \a^{(1-2n)/2n}   \over d-1}
\sim \eta^{1\over d-1},~~~~~~~~~~~~~~~~~~ \eta>0,
\label{332}
\eea
where $d\eta= dt/a= d\ti t/\ti a$. The two asymptotic branches of $\ti a(\eta)$ are thus simply related by a time reversal transformation, as appropriate to a ``self-dual" background \cite{10} in the Einstein frame.
The Einstein scale-factor $\ti a(\eta)$, on the other hand, represents the ``pump field" determining asymptotically the spectrum of scalar and tensor metric perturbations \cite{11} (as also illustrated in the subsequent sections). All the solutions of the considered class are thus expected to lead to the same asymptotic spectrum of amplified metric fluctuations, irrespectively of the values of $\a$ and $n$. For the analytical and numerical computations of this paper it will thus be enough to use the particularly simple case $n=1$, $\a=d$, $m^2=V_0$ of the general class (\ref{325}), corresponding to the (non-local) two-loop potential $V=-V_0 e^{4 \vpb}$. In such a case the exact   solution, for $d=3$ spatial dimensions, can be explicitly written as \cite{5} 
\bea
&&
a(\tau) =  \biggl[ \tau + \sqrt{\tau^2 + 1}\biggr]^{1/\sqrt{3}},
~~~~~~~~
 \vpb =\vp_{0} - \frac{1}{2} \ln{ ( 1 + \tau^2)} , \nonumber\\
&&
\tau=t/t_0, ~~~~~~~~~~~~~~~~~~~~~~~~~~~~~~~
t_0^{-1}= e^{2\vp_0} \sqrt{ V_0},
\label{333}
\eea
where (without loss of generality) we have set $a=1$ at the bounce, occurring at $t=0$, and where $\vp_0$ is an integration constant. Such solution is characterized by the two relevant parameters,  $t_0^{-1}$ and $e^{{\vp}_0}$, corresponding (up to numerical factors) to the Hubble parameter and string coupling at the bounce. 

For later applications it is now convenient to present also the homogeneous limit of the E-frame equations (\ref{413}), (\ref{415}). Considering a  homogeneous, isotropic and spatially flat E-frame background we obtain, from Eq. (\ref{44}) and (\ref{414}), 
\beq
-\vpb={c\over 2} \ti \vp + d \ln \ti a, ~~~~~~~
\sqrt{\pa \ti \vp^2} ~ J_1 e^{c\ti \vp/2} =- {\pa V\over \pa  \vpb},
\label{419}
\eeq
where we have absorbed, as before, the total, finite volume (now in Planck units) of the compact spatial sections. 
The $(00)$, $(ij)$ components  of the Einstein equations (\ref{413}), and the dilaton equation (\ref{415}), thus become, respectively (in units $2 \lp^{d-1}=1$): 
\bea
&&
d(d-1) \ti H^2 = \ti \r +{1\over 2} {\dot {\ti \vp}}^2+e^{c\ti \vp}V,
\nonumber\\
&&
2(d-1)\dot {\ti H} + d(d-1) \ti H^2 =-\ti p -\left({1\over 2} {\dot {\ti \vp}}^2-e^{c\ti \vp}
V\right)- e^{c\ti \vp}{\pa V\over \pa
\vpb}, \nonumber \\
 &&
\ddot {\ti \varphi} +d\ti H \dot{\ti  \vp} +{1\over 2} \ti \sg 
+{c\over2} (\ti \r-d\ti p)
+e^{c\ti \vp}
\left(cV-{c\over 2}{\pa V\over \pa \vpb}
\right) =0, 
\label{420}
\eea
where the dot now denotes differentiation with respect to the E-frame cosmic time $\ti t$. 
One can easily check that such equations can also be directly obtained  from the corresponding S-frame equations (\ref{319})-(\ref{321}), by performing the canonical rescaling
\bea
&&
a \to \ti a=a e^{- \vp/(d-1)}, ~~~~~~~~ 
dt \to d \ti t=dt  e^{- \vp/(d-1)}, ~~~~~~~
\vp \to \ti \vp=c \vp, 
\nonumber \\
&&
 \r \to \ti \r=\r e^{{d+1\over d-1} \vp},  ~~~~~~~~~~~~~
p \to \ti p=p e^{{d+1\over d-1} \vp},  ~~~~~~~~~~~~~~~
 \sg \to \ti \sg={\sg\over c} e^{{d+1\over d-1} \vp},
\label{421}
\eea
connecting the two frames. 
The combination of the three equations (\ref{420}), in particular, leads to the conservation equation 
\beq
\dot {\ti \r} +d\ti H(\ti \r+\ti p) -{c\over2}(\ti \r-d\ti p) \dot {\ti \vp}
-{1\over 2} \ti \sg \dot {\ti  \vp}=0, 
\label{422}
\eeq
and to the useful identity:
\beq
2(d-1)\dot{\ti  H} +\ti \r+\ti p+ {\dot {\ti \vp}}^2+{\pa V\over \pa \vpb}
e^{c\ti \vp}=0. 
\label{423}
\eeq

It is important to compare these equations with those obtained from an action identical to (\ref{49}), but with a {\em local} potential $\ti V = e^{c \ti \vp} V(\ti \phi)$.  The $(00)$ equation and the conservation
equation are the same,  but the other equations are different. One obtains, for a local (S-frame) potential $V(\vp)$, 
\bea
&&
d(d-1) \ti H^2 = \ti \r +{1\over 2} {\dot {\ti  \vp}}^2+e^{c\ti \vp}V,
\nonumber\\
&&
2(d-1)\dot {\ti H}+d(d-1) \ti H^2 =-\ti p -\left({1\over 2} {\dot {\ti \vp}}^2-e^{c\ti \vp}V\right), \nonumber \\
 &&
\ddot {\ti \varphi} +d\ti H \dot {\ti \vp} +{1\over 2} \ti \sg 
+{c\over2}  (\ti \r-d\ti p)
+e^{c\ti \vp}
\left(cV+ {\pa V\over \pa \ti \vp}\right) =0.
\label{424}
\eea
The combination of the first two equations, in particular, leads to the usual identity
\beq
2(d-1)\dot {\ti H} +\ti \r+\ti p+ {\dot{\ti \vp}}^2=0,
\label{425}
\eeq
so that the behaviour of $\dot {\ti H}$ is completely determined by ${\dot {\ti \vp}}^2$ and by the fluid variables. For the background solutions considered in this paper this is no longer true, because of the explicit contribution of the potential to Eq. (\ref{423}). 

The (E-frame) background equations (\ref{420}) will be applied, in the following sections, to discuss the evolution of scalar metric perturbations, and to compute their final spectral distribution. 
In $d=3$ spatial dimensions, and in terms of the frame-invariant conformal time parameter $\eta$, they can be written explicitly as
\bea 
&&
6  {\ti {\cal H}}^2 = \ti \r \ti a^2 +{1\over 2} {\ti  \vp}^{\prime 2}+e^{\ti \vp}V {\ti a}^2,
\label{426}\\
&&
4 {\ti {\cal H}}'+2 {\ti {\cal H}}^2 =-\ti p \ti a^2 -\left({1\over 2} {\ti  \vp}^{\prime 2}-e^{\ti \vp}
V{\ti a}^2\right)- e^{\ti \vp}{\pa V\over \pa \vpb}{\ti a}^2, \label{427} \\
&&
 \ti \varphi^{\prime \prime} +2 \ti {\cal H} {\ti \vp}'  +{1\over 2} \ti \sg {\ti a}^2
+{1\over 2}(\ti \r -3\ti p){\ti a}^2 
+e^{\ti \vp}
\left(V-{1\over  2}{\pa V\over \pa \vpb}\right){\ti a}^2
=0,
\label{428}
\eea
where $\ti {\cal H}={\ti a}'/\ti a$, and the prime denotes differentiation with respect to $\eta$. 
The  conservation equation and the identity (\ref{423}) read, respectively,
\bea
&&
\ti \r' +3 \ti {\cal H}(\ti \r+\ti p) -{1\over2} \ti \vp' (\ti \r-3\ti p)-{1\over2} \ti \sg \ti \vp' =0,
\label{429}\\
&&
4( \ti {\cal H}' - {\ti {\cal H}}^2) + {\ti  \vp}^{\prime 2}+(\ti \r+\ti p) {\ti a}^2 +e^{\ti \vp} {\pa V\over \pa \vpb}{\ti a}^2=0.
\label{430}
\eea

It is finally worth noticing, even if not essential for the understanding of our subsequent discussion on the matching of perturbations, that regular bouncing solutions are possible even in the presence of matter sources. Working with the S-frame equations we may consider, in particular, the special case in which  $\pa V/\pa \vpb= 2V$, and in which  the homogeneous equations (\ref{319})-(\ref{321}) can be reduced to quadratures for a wide class of equations of state satisfying the condition that the ratio $\pb/\rb$ is an integrable function of the time-coordinate $x$, related to cosmic time by \cite{4}
\beq
2 dx= L\rb dt, ~~~~~~~~~~~~ L= {\rm const}.
\label{334}
\eeq
We can take, for instance,
\beq
V=-V_0 e^{2 \vpb}, ~~~~~~~~~~
\pb= \ga \rb, ~~~~~~~~~~ \sb = \ga_0 \rb,
\label{335}
\eeq
with $V_0$, $\ga$, $\ga_0$ constants. In such a case, assuming that $\ga \not= \ga_0/2$ and that $\ga_0/2\not=1$, a first integration and a combination of the above equations leads to \cite{12}
\beq
\vpb' =-2 \left(1-{\ga_0\over 2}\right){x+x_0\over D}, ~~~~~~~~
{a'\over a}= 2\left(\ga-{\ga_0\over 2}\right) {x+x_1\over D}, 
\label{336}
\eeq
where
\beq
D(x)=L^2 e^{-\vpb} \rb = \left(1-{\ga_0\over 2}\right)^2(x+x_0)^2
-d\left(\ga-{\ga_0\over 2}\right)^2 (x+x_1)^2 +L^2V_0,
\label{337}
\eeq
and $x_1,x_0$ are integration constants. A second integration may then provide regular solutions for $a$ and $\vpb$ if $D(x)$ has no real zeros, namely if the discriminant $\Da^2$ of the quadratic form $D(x)$ satisfies  $\Da^2<0$. This is possible {\em provided} $V_0>0$, i.e. for $V(\vpb)<0$. In the simple case $\ga_0=0$ we get, for instance,
\beq
\Da^2= d\ga^2(x_1-x_0)^2 +L^2V_0(d \ga^2 -1),
\label{338}
\eeq
so that the condition $\Da^2<0$ can be implemented for a set of reasonable equations of state, characterized by $d\ga^2 <1$. 

In the special case $\sg=0$ and  $p=0$ we are lead, with a first integration, not to eqs. (\ref{336}), but to the following set of equations \cite{12},
\beq
\vpb' =-2 {x+x_0\over D}, ~~~~~~~~
{a'\over a}=  {2x_1\over D}, ~~~~~~~~
D=(x+x_0)^2-dx_1^2+L^2V_0.
\label{339}
\eeq
Regular solutions are thus allowed for $\Da^2=dx_1^2-L^2V_0<0$.
The choice $x_1=0$, $x_0=0$, leads to the particularly simple exact  solution describing a trivially flat (String-frame) space-time, with the dilaton performing a (time-symmetric) bouncing evolution sustained by the presence of a constant energy density of the (dust) matter sources:
\beq
a=a_0, ~~~~~~~ \r=\r_0, ~~~~~~~ p=0, ~~~~~ ~~e^\vp={e^{\vp_0}\over 1+(t/t_0)^2},
\label{340}
\eeq
where $a_0$, $\r_0$ are integration constants, and
\beq
 e^{\vp_0}\r_0={4\over t_0^2}=V_0e^{2\vp_0}.
\label{341}
\eeq

It should be noted that such solution acquires a less trivial representation in the Einstein frame where, by using the appropriate transformations (\ref{421}) one finds, in conformal time,
\bea
&&
a_E(\eta)= a_0e^{-\vp_0/(d-1)} \left(1+{\eta^2\over \eta_0^2}\right)^{1\over d-1}, ~~~~~~~~~
e^\vp= e^{\vp_0} \left(1+{\eta^2\over \eta_0^2}\right)^{-1}, \nonumber \\
&&
\r_E(\eta)=\r_0 a_0^d \left(1+{\eta^2\over \eta_0^2}\right)^{-{d\over d-1}},
\label{342}
\eea
where $\eta_0=t_0/a_0$. In this frame the metric smoothly evolves from accelerated contraction to decelerated expansion, sustained by the dilaton and by a non-trivial, ``bell-like" evolution of the energy density of the pressureless fluid. The two asymptotic branches of such a ``self-dual" solution are simply related, in the Einstein frame,  by a time-reversal transformation, like in the previous class of vacuum solutions 
(\ref{332}). 
This simple background, together with a more general class of solutions, will be used in a forthcoming paper \cite{8} to discuss the spectrum of metric perturbations in the presence of perfect fluid sources.

\section{Perturbation equations in the general case}
\label{Sec4}
\setcounter{equation}{0}
\subsection{Perturbation equations in the Einstein frame}
\label{Sec4.1}

In this section we will perturb the E-frame background equations presented in subsection \ref{Sec3.2}, by expanding the background fields around the homegeneous configurations of subsection \ref{Sec3.3},  in order to obtain the linearized equations governing the evolution of the (first order) gravi-dilaton and matter fluctuations. Note that, from now on, we will drop the tilde for the sake of a simpler notation, but all variables appearing in the subsequent equations are to be interpreted (unless otherwise stated) as E-frame variables. 

The computation of the tensor (transverse and traceless) part of  the metric perturbations can be easily performed in the syncronous gauge, by setting 
\bea
&& 
g_{00}=1, ~~~~~~~~~~~~~~~~g_{0i}=0, ~~~~~~~~ g_{ij}=-a^2\da_{ij}, \nonumber \\
&&
\da g_{00}=0= \da g_{i0}, ~~~~~\da g_{ij}=h_{ij}, ~~~~~ g^{ij}h_{ij}=0=\pa_j h_i\,^j ,
\label{51}
\eea
while keeping the dilaton and the matter sources unperturbed, $\da \vp=0$, $\da T_{\mu\nu}=0= \da \sg$. We then find that all the perturbed terms of Eqs. (\ref{413}), (\ref{415}) are vanishing, except
\beq
\da R_i~^j= -{1\over 2}\Box h_i\,^j \equiv -{1\over 2} \left( \ddot  h_i\,^j
+dH \dot  h_i\,^j -{\nabla^2\over a^2}  h_i\,^j\right).
\label{52}
\eeq
For each polarization mode $h$ we thus recover the usual (E-frame) evolution equation, which can be written in Fourier space (using conformal time and in $d=3$) as
\beq
 h_k'' + 2 {\cal H}_{e}  h_k'+k^2h_k=0.
\label{53}
\eeq 
The associated canonical variable \cite{11}
\beq 
\mu = {\Mp \over \sqrt 2} a h, ~~~~~~~
\mu'' -\left( \nabla^2 +{a''\over a}\right)=0,
\label{54}
\eeq
diagonalizes the kinetic term in the effective action for tensor metric perturbations, has canonical dimensions and can be correctly normalized to an initial spectrum of quantum vacuum fluctuations ($|\mu_k| \to |2 k|^{-1/2}$ for $ \eta \to -\infty$). 
For a thorough discussion of tensor perturbations see \cite{7}.

To determine the evolution of scalar perturbations in a conformally flat background, $g_{\mu\nu}={\rm diag}~ a^2 (1, -1, -1, -1, \dots)$, we will keep using the conformal time coordinate, and we will work in the so-called ``uniform-dilaton" gauge \cite{7}, by setting
\bea
&&
\da \vp=0, ~~~~~~~~~~ \da g_{00}= 2 a^2 \phi, ~~~~~~~~~ \da g_{ij} =2 a^2 (\psi \da_{ij} -\pa_i\pa_jE), \nonumber\\
&&
\da T_0~^0=\da \r, ~~~~~~  \da T_i~^j=-\da p \da_i^j, ~~~~~~ \da
T_i~^0= (\r +p)\pa_i u,
\label{55}
\eea
where $u$ is the usual velocity potential for the fluid perturbations \cite{11}. We find, in this gauge, 
\beq
\da \ga_\mu^\nu=0, ~~~~~~~~~~
\da (e^{-\vpb})=0, ~~~~~~~~~~
\da I_1=0.
\label{56}
\eeq
The perturbation of the Einstein equations (\ref{413}) leads then to
\beq
\delta G_{\mu}^{\nu} = {1\over 2}\da T_\mu^\nu+ {1\over 2}
\delta g^{\nu\alpha}\partial_{\mu} \vp \pa_\a \vp 
-{1\over 4} \da_\mu^\nu\left(\delta g^{\alpha\beta}
\partial_{\alpha}\vp \partial_{\beta} \vp \right)
+{1\over2} e^\vp \ga_\mu^\nu {\pa V\over \pa \vpb} \phi, 
\label{57}
\eeq
where the last term of this equation has been computed using already the homogeneity assumption $\vp=\vp(\eta)$. 

Let us consider the above equation for the case $d=3$. The $(i \not=j)$ component gives
\beq
E'' + 2 {\cal H} E' + \psi - \phi=0.
\label{58}
\eeq
The $(i0)$ component gives
\beq
\psi'+{\cal H}\phi={1\over 4}(\r+p)a^2 u.
\label{59}
\eeq
The $(00)$ component gives
\beq
\nabla^2(\psi+{\cal H} E')-3 {\cal H}(\psi'+{\cal H}\phi)={1\over 4}
a^2 \da \r -{1\over 4}\vp'^2 \phi.
\label{510}
\eeq
The $(i=j)$ component gives
\beq
(2 {\cal H}'+{\cal H}^2)\phi +{\cal H}\phi' +\psi''+2 {\cal H} \psi'={1\over 4}
a^2 \da p -{1\over 4}\vp'^2 \phi -{1\over 4} e^\vp {\pa V\over \pa
\vpb} a^2 \phi
\label{511}
\eeq
(we have also used the $(i \not=j)$ condition, i.e Eq. (\ref{58})). 

Consider now the limit in which the fluid sources (and fluid  perturbations) are absent (or negligible). 
In that case, the first two equations are reduced to
\begin{eqnarray}
&& E'' + 2 {\cal H} E' + \psi - \phi=0,
\label{512}\\
&&  \psi' + {\cal H} \phi=0,
\label{513}
\eea
Eq. (\ref{510}), using the momentum constraint (\ref{513}) and the background equation (\ref{426}), can be rewritten as
\beq
2\nabla^2(\psi+{\cal H} E')-6{\cal H}\psi'-e^\vp V a^2 \phi=0.
\label{514}
\eeq
Eq. (\ref{511}),  using the 
background equation (\ref{426}) for $\vp'^2$, and the 
background equation (\ref{427}) for ${\pa V/ \pa\vpb} $, can be rewritten as 
\beq
 \psi'' + 2 {\cal H} \psi' + {\cal H}\phi' + ({\cal H}' + 2{\cal
H}^2) \phi =0.
\label{515}
\eeq
It can be easily checked that the four above equations exactly correspond to the E-frame transformed version of the perturbation equations computed in the S-frame in the absence of fluid sources, already presented in our previous paper \cite{7}. 

In the presence of fluid sources, the system of equation 
(\ref{58})-(\ref{511}) can be completed 
by perturbing the matter conservation equation (\ref{418}).  For the application of this paper we will limit ourselves to the case $\sg=0$, $ \da \sg=0$. The time component of the perturbed equation gives then 
\beq
\da \r' -(\r+p) \nabla^2 u+ (\r+p) (\nabla^2 E' -3 \psi')
+3 {\cal H}(\da \r +\da p) = {1\over 2} \vp' (\da \r - 3 \da p),
\label{516}
\eeq
while the space component gives
\beq
u' +4 {\cal H} u +{\r'+p'\over \r +p}u -\phi ={\da p\over \r +p}.
\label{517}
\eeq

\subsection{Gauge-invariant  scalar perturbations}
\label{Sec4.2}

The perturbation equations written in the previous subsection have been obtained using the particular gauge (\ref{55}), and are not invariant under the infinitesimal coordinate transformations
\beq
\eta \to \eta + \ep^0(\eta, x^i), ~~~~~~~~~~~
x^i \to x^i + \pa^i \ep (\eta, x^i), 
\label{518}
\eeq
still preserving the scalar nature of the metric fluctuations. However, even within the more general decomposition of dilaton and scalar metric perturbations, i.e. 
\bea
&&
\da \vp = \chi, ~~~~~~~~~~~~~~~~~~\da g_{i0}= a^2 \pa_i B, 
\nonumber \\
&&
\da g_{00}= 2 a^2 \vp, ~~~~~~~~~~~ \da g_{ij} =2 a^2 (\psi \da_{ij} -\pa_i\pa_jE),
\label{519}
\eea
and for any given choice of gauge for the variables $ \{\chi, \phi, \psi, E, B, \da \r, \da p, u \}$, it is always possible to define an independent set of gauge-invariant perturbations. For instance \cite{11}, the set of variables $ \{\Phi, \Psi, X,  {\cal E}, P, W \}$ defined by
\bea
&&
\Phi= \phi +{1\over a} \left[ (B-E') a\right]', ~~~~~~~~
\Psi=\psi -{\cal H}(B-E'), \nonumber \\
&&
X= \chi + \vp' (B-E'), ~~~~~~~~~~~~
{\cal E}= \da \r + \r' (B-E'), 
\nonumber \\
&&
P= \da p + p' (B-E'), ~~~~~~~~~~~~
W= u + B- E',
\label{520}
\eea
is invariant under the infinitesimal transformation (\ref{518}). The gauge invariant Bardeen potentials $\Psi$ and $\Phi$, in particular, coincide with the scalar components of the metric fluctuations in the so-called longitudinal gauge $E=0=B$ \cite{11}. We also introduce, for later use, another useful gauge invariant combination,
\beq
{\cal R}= -\psi - {{\cal H}\over \vp'}  \chi= -\Psi-{{\cal H}\over \vp'} X,
\label{521}
\eeq
representing (in the absence of fluid sources) the perturbation of the spatial part of the scalar curvature \cite{13} with respect to comoving hypersurfaces, i.e., spacelike $3$-d surfaces characterized by a uniform dilaton distribution. 
In our case, the perturbation variables associated to the uniform-dilaton gauge can be expressed in terms of the variables (\ref{520}) as follows:
\bea
&&
E'= -{X\over \vp'}, ~~~~~~~~~~~~ \psi=\Psi +{ {\cal H} \over \vp'} X,
~~~~~~~~
\phi= \Phi -{ {\cal H} \over \vp'} X - { X'\over \vp'} +{\vp''\over \vp'} X,
\nonumber \\
&&
\da \r = {\cal E} - {\r'\over \vp'} X, ~~~~~~
\da p = P - {p'\over \vp'} X, ~~~~~~~~
u=W-{X\over \vp'} .
\label{522}
\eea
By inserting these defintions into the previous set of equations (\ref{58})-(\ref{511}), (\ref{516})-(\ref{517}), we can thus obtain a linearized set of perturbation equations written in a fully gauge-invariant form. In particular, Eq. (\ref{58}) becomes
\beq
\Phi=\Psi.
\label{523}
\eeq
Eq. (\ref{59}) becomes
\beq
\Psi'+{\cal H}\Psi+{X\over \vp'}({\cal H}'-{\cal H}^2)=
{1\over 4}(\r+p)a^2 \left(W-{X\over \vp'}\right).
\label{524}
\eeq
Eq. (\ref{510}) becomes
\bea
&&
\nabla^2\Psi-\left(3 {\cal H}^2-{\vp'^2\over 4}\right)\Psi
-3{\cal H} \Psi'+ {\cal H}{X\over \vp'}\left(3 {\cal H}^2
-3{\cal H}'-{\vp'^2\over 4}\right) \nonumber\\
&&
-{1\over 4}X'\vp'+{1\over 4}X\vp''
-{1\over 4}
a^2 \left({\cal E} -\r' {X\over \vp'}\right)=0.
\label{525}
\eea
Eq. (\ref{511}) (using the identity (\ref{430})) gives
\bea
&&
\Psi''+3{\cal H}\Psi'+\left[{\cal H}'+ 2 {\cal H}^2-{1\over 4}
(\r+p)a^2\right] \Psi
+{X'\over \vp'}\left[{\cal H}'- {\cal H}^2+{1\over 4}
(\r+p)a^2\right]  \nonumber\\
&&
+{X\over \vp'}\left[{\cal H}''- 2{\cal H}^3+{{\cal H}\over 4}
(\r+p)a^2 +{\vp''\over \vp'}\left({\cal H}^2-{\cal H}'-{1\over 4}
(\r+p)a^2\right)\right] \nonumber\\
&&
={1\over 4} a^2\left(P-p'{X\over \vp'}\right)
\label{526}. 
\eea
Finally, Eq. (\ref{516}), using the background conservation
equation (\ref{429}) (with $\sg=0$), can be rewritten as
\beq
{\cal E}'- (\r +p) \nabla^2 U - 3 (\r +p) \Psi' + 3 {\cal H}({\cal E}+P)
-{1\over 2} \vp'({\cal E}-3P) -{1\over 2} X' (\r -3p)=0.
\label{527}
\eeq
Eq. (\ref{517}), using again the conservation
equation, can be rewritten as
\beq
W'+ {\cal H} W - \Psi +{1\over \r+p}\left[p' W - P+{1\over 2}\vp' (\r-3p)(W-
{X\over \vp'})\right]=0.
\label{528}
\eeq

\subsection{Relation to String-frame scalar perturbations}
\label{Sec4.3}

 In order to help the reader to make contact with previous results \cite{7}, it is useful to provide explicit relations between the E-frame and the 
S-frame perturbation variables. Unlike tensor fluctuations, scalar fluctuations are indeed neither gauge- nor frame-independent. Of course, since infinitesimal coordinate transformations have the same form in both frames, gauge-invariant perturbations can be defined in the same way in each frame, so that Eqs. (\ref{520}), (\ref{521}) can also be interpreted as the definition of gauge-invariant variables in the S-frame, provided the background and perturbations variables are referred to the S-frame metric, dilaton and fluid sources. 

The relation between a generic set of scalar perturbations in the E-frame (see for instance Eq. (\ref{519}), completed by the fluid perturbations of Eq. (\ref{55})), and the corresponding set in the S-frame, can be easily obtained by recalling the transformation (\ref{421}):
\bea
&&
g_{\mu\nu}^E= g_{\mu\nu}^s e^{-\vp}, ~~~~~~~~~
a_E= a_s e^{-\vp/2}, ~~~~~~~~~
\H_E= \H_s - \vp'/2.
\nonumber\\
&&
\vp_E= \vp_s=\vp, ~~~~~~~~~
(T_E)_\mu^\nu = (T_s)_\mu^\nu~ e^{2 \vp}.
\label{529}
\eea
We are considering the case $d=3$, and we have appended to our variables the labels $E$ and $s$ for an easier distinction of the two frames.  It follows that 
\beq
\da g_{\mu\nu}^E = \da g_{\mu\nu}^s e ^{-\vp} - \chi~ \da g_{\mu\nu}^E,
~~~~~~
(\da T_E)_\mu^\nu= e^{2 \vp} (\da T_s)_\mu^\nu+ 2 \chi 
(\da T_E)_\mu^\nu, 
\label{530}
\eeq
from which, using the definitions (\ref{519}), (\ref{55}) (valid in both frames), 
\bea
&&
\phi_E= \phi_s -{\chi \over 2}, ~~~~~
\psi_E= \psi_s +{\chi \over 2}, ~~~~~
E_E=E_s, ~~~~~ B_E=B_s, ~~~~~ \chi_E= \chi_s,
\nonumber\\
&&
\da \r_E = \da \r_s e^{2\vp}+ 2 \chi \r_E, ~~~~~~
\da p_E = \da p_s e^{2\vp}+ 2 \chi p_E, ~~~~~~
u_E=u_s. 
\label{531}
\eea
For our uniform-dilaton gauge, in particular, $\chi_E=\chi_s=0$, so that all the metric perturbation variables coincide in the two frames, while the fluid perturbation variables only differ by an overall rescaling factor. 

For what concerns the gauge-invariant variables, using Eqs. (\ref{529}), (\ref{531}) we obtain, from the definitions (\ref{520}) and (\ref{521}):
\bea
&&
\Phi_E= \Phi_s - {\chi \over 2} -{\vp' \over 2}(B-E')
=\Phi_s -{1\over 2} X_s, ~~~~~~~~~~
X_E= X_s,
\nonumber\\
&&
\Psi_E= \Psi_s + {\chi \over 2} +{\vp' \over 2}(B-E')
=\Psi_s +{1\over 2} X_s, ~~~~~~
~~~~ \R_E= \R_s. 
\label{532}
\eea
The (gauge-invariant) scalar field and curvature perturbations $X$ and $\R$ are thus frame-independent, together  with the (gauge-invariant) potential $W$, and the combination $\Psi+\Phi$. The two Bardeen potentials, on the contrary, are not separately frame-independent.

\section{Solutions for scalar perturbations} 
\label{Sec5}
\setcounter{equation}{0}
\subsection{General analytic considerations}
\label{Sec5.1}
In this section we will compute the spectrum of scalar perturbations associated with the regular gravi-dilaton background (\ref{333}). We will apply the equations obtained in the previous section by setting everywhere $\r=p=0$, and $\da \r=\da p=u=0$, as we shall consider a pure gravi-dilaton system with no fluid sources. 

We are interested, in particular, in studying the evolution and the final spectrum of the Bardeen potential $\Psi$ and of the curvature perturbation $\cal R$. We thus need the decoupled equations describing the evolution of such variables. Let us first note, to this purpose, that a decoupled equation for $\Psi$ can be obtained by expressing $X$ and its first derivative from Eq. (\ref{524}), and using such result to eliminate $X$ and $X'$ from Eq. (\ref{525}). The resulting equation can be written in the form
\beq
\Psi'' +2 {\xi'\over \xi} \Psi' +2\left(\H' -\H^2 +\H {\xi'\over \xi}\right)\Psi 
-4{(\H^2-\H')\over \vp'^2} \nabla^2 \Psi=0,
\label{61}
\eeq
where
\beq
\xi=a/(\H^2-\H')^{1/2}
\label{62}
\eeq

It can be easily checked that, for a {\em local} potential, one has $4 
(\H^2-\H')= \vp'^2$ (see Eq. (\ref{425})), and one recovers from (\ref{61}) the standard evolution of the Bardeen potential for a minimally coupled scalar field (see for instance \cite{11}). In our case, the first derivative term in eq. (\ref{61}) can be eliminated by introducing the variable $U= \xi \Psi$, and the equation can be rewritten in ``pseudo-canonical" form as
\beq
 U'' -\left (c_s^2 \nabla^2 + { Z''\over Z} \right) U=0,
\label{63}
\eeq
where, using the background identity (\ref{430}),
\bea
&&
c_s^2={4\over \vp'^2} (\H^2-\H')=
 1 + {a^2 \over \vp'^2} {\pa V\over \pa \vpb} e^\vp, \nonumber\\
&&
Z= {\H \over 2a (\H^2-\H')^{1/2}}={\H\over a \vp' c_s}.
\label{64}
\eea

For what concerns the equation for curvature perturbations we may note that, in our gauge, $\da \vp=0$, so that $\R= -\psi$ (from the definition (\ref{521})). The decoupled equation for $\R$ can thus be directly obtained from the decoupled equation for $\psi$ by combining Eqs.
(\ref{512})-(\ref{515}), and by using the background equations 
(\ref{426})-(\ref{428}). The final result is 
\beq
\R'' + 2 {z'\over z} \R' -c_s^2 \nabla^2 \R=0,
\label{65}
\eeq
where
\beq
z=a\vp'/\H,
\label{66}
\eeq
and the first derivative term can be eliminated by introducing
the variable $v= z \R $, which satisfies the standard equation
\beq
v'' - \left(c_s^2 \nabla^2 +{z''\over z}\right)v=0, ~~~~~~~~
v= z \R  =- a \left( X+ {\vp'\over \H} \Psi\right).
\label{67}
\eeq

It is important to stress that $v$, in the case of a local potential,
exactly represents (modulo a sign, and in units $\Mp^2=2$), the
gauge-invariant canonical variable diagonalizing the action for scalar
perturbations \cite{11}. In our case Eq. (\ref{67}) is different from the
canonical one because of the effective ``speed coefficient"
$c_s^2\not= 1$. However, in our background (\ref{333}), $c_s^2-1$
goes rapidly to zero at large negative times so that, even in our case,
$v$ is the appropriate variable for the asymptotic normalization of
scalar perturbations to a vacuum fluctuation spectrum, namely
\beq |v_k| =(2 k)^{-1/2}
\label{68}
\eeq
for $\eta \ra -\infty$. The normalization of $v_k$ then fixes the
asymptotic normalization of the $\R_k$ modes, through the definition
$\R=v/z$. 

The normalization of $\R$, in its turn, controls the quantum
normalization of the Bardeen potential, once a relation between $\R_k$
and $\Psi_k$ is given. Such a relation can be easily obtained again
working in the uniform dilaton gauge, where $\R=-\psi$ and
$\Psi=\psi+\H E'$. By eliminating $\phi$ in Eq. (\ref{514}) through Eq.
(\ref{513}), and $Va^2e^\vp$ through the background equation
(\ref{426}), one is lead to the relation
\beq
\R'=-\psi'=-{4\H \over \vp'^2} \nabla^2 \Psi,
\label{69}
\eeq
which holds exactly for all modes and at all times. For the asymptotic
normalization at $\eta \ra -\infty$, where all modes are oscillating
outside the horizon ($\R_k \sim \Psi_k \sim e^{-ik\eta}$), we then
find, in particular,
\beq
\left|\Psi_k\right| ={3\over 2} {\left| \R_k \right|\over \left| k\eta
\right|},~~~~~~~~~~ \left| k\eta\right| \gg 1.
\label{610}
\eeq
It follows that, for what concerns the $k$-dependence of the initial
fluctuations, normalized to a vacuum fluctuation spectrum, one has
$\R_k \sim v_k \sim k^{-1/2}$, while $\Psi_k \sim U_k \sim k^{-3/2}$. 

It should be noted, finally, that in the case of a local potential
$c_s^2=1$, so that $Z=z^{-1}$ (see Eqs. (\ref{64}), (\ref{66})), and the
variables $U$ and $v$ satisfy the same canonical equation with
``duality-related" pump fileds. Such a duality relation is broken, in
general, for a non-local potential, but it becomes formally restored
asymptotically, at large enough times. 

\subsection{Analytic estimates using different matching prescriptions}
\label{Sec5.2}

We shall now apply the above equations to the computation of the
scalar perturbation spectrum associated with the background (\ref{333}). 

We start noticing that for such background, sufficiently far from the
bounce, the dilaton potential becomes negligible, and one recovers the
usual vacuum solution of minimal pre-big bang models \cite{1}:
\bea
&&
a \sim (-\eta)^{1/2}, ~~~~~~ \vp \sim - \sqrt 3 \ln (-\eta), 
~~~~~~ \eta \ra -\infty, \label{611} \\
&&
a \sim (\eta)^{1/2}, ~~~~~~~~ \vp \sim  \sqrt 3 \ln (\eta), 
~~~~~~~~~~~ \eta \ra +\infty,
\label{612}
\eea
It follows that, asymptotically, $c_s^2=1$, $z''/z=a''/a$, and Eq.
(\ref{67}) for $v$ (or for $\R$, Eq. (\ref{65})), exactly reduces to Eq. 
(\ref{54}) for $\mu$ (or for $h$, Eq. (\ref{53})). We thus expect, at least
for the low-energy modes exiting (and then re-entering) the horizon
far enough from the bouncing region, that the curvature fluctuations
$\R_k$ should be amplified with a scalar spectrum identical to that of
tensor perturbations. 

For such low-energy modes we know that the initial solutions,
normalized to the quantum fluctuations of the vacuum, are given in
terms of the Hankel functions of second kind and index zero, $h_k \sim
\R_k \sim (|\eta^{1/2}|/a) H_0^2(|k\eta|)$, which describe a
logarithmic growth of the fluctuations outside the horizon. In the case
of tensor modes, on the other hand, there seems to exist an almost unanimous
consensus on the fact that the final spectrum should be computed by
assuming a smooth evolution of the metric fluctuation $h$ across the
bounce. In the background (\ref{611}), (\ref{612}) this leads to a
low-energy spectrum which, after all modes re-enter the horizon 
($\eta \ra +\infty$), can be parametrized as 
\beq
|\da_h| = k^{3/2} |h_k|={k\over k_1}\left [A_1 +A_2 \ln
\left(k\over k_1\right)\right], 
\label{613}
\eeq
where $k_1$ is the high-energy cut-off scale, and $A_1,A_2$ are
constant (and $k$-independent) dimensionless coefficients depending
on the initial normalization.  After the bounce, but still outside the
horizon ($k\eta \ll 1$), the spectrum takes the form
\beq
|\da_h| = k^{3/2} |h_k|=\left(k\over k_1\right)^{3/2}
\left [B_1 +B_2 \ln \left(k_1\eta \right)\right].  
\label{614}
\eeq
For the scalar modes $\R_k$, of energy low enough to be unaffected by
the specific kinematics of the bouncing region, we would thus expect a
similar spectral behaviour, namely
\beq
|\da_\R| = k^{3/2} |\R_k|={k\over k_1}\left [\a_1 +\a_2 \ln
\left(k\over k_1\right)\right]  
\label{615}
\eeq
inside the horizon, and 
\beq
|\da_\R| = k^{3/2} |\R_k|=\left(k\over k_1\right)^{3/2}
\left [\b_1 +\b_2 \ln \left(k_1\eta \right)\right]   
\label{616}
\eeq
outside the horizon. Given $\da_\R$, the corresponding Bardeen
spectrum after the bounce is then determined by Eqs. (\ref{69}),
(\ref{610}) outside and inside the horizon, respectively, as
\beq
|\da_\Psi| \sim {|\da_\R|\over |k\eta|^2}, ~~~~ k\eta \ll 1, ~~~~~~~~
|\da_\Psi| \sim {|\da_\R|\over |k\eta|}, ~~~~ k\eta \gg 1.
\label{617}
\eeq

The above expression for $\da_\R$ and $\da_\Psi$ will be fully
confirmed by a detailed numerical computation, as we shall see in the
next subsection. Before presenting numerical results, however, we want 
 to stress with an explicit example how the above spectrum 
is crucially dependent on the assumption that the variable $\R$ goes
smoothly through the bounce, and how the alternative assumption of a
smooth behaviour of $\Psi$ would lead instead to different spectra, in
disagreement with subsequent numerical computations. 

Let us suppose, for our illustrative purpose, that in the absence of a
full and explicit knowledge of the background dynamics near the
bouncing transition we want to compute the scalar spectrum by
performing a matching of the perturbation variables, from
$\eta=-\eta_1$ to $\eta=+\eta_1$, where $\eta_1$ is  typically the
transition time scale. Let us also suppose that, outside the transition
regime (i.e. for $\eta \leq -\eta_1$ and $\eta \geq \eta_1$), the
contribution of the dilaton potential becomes negligible, and the
background evolution can be correctly approximated by the aymptotic
solutions (\ref{611}), (\ref{612}). We shall put, in
particular,
\bea
&&
a=\left(-\eta\over \eta_1\right)^{1/2}, ~~~~
\H={1\over 2 \eta}, ~~~~
\vp' =-{\sqrt 3 \over \eta}, ~~~~
z=- 2 \sqrt 3 a, ~~~~~~~
\eta \leq -\eta_1,
\label{618}\\
&&
a=\left(\eta\over \eta_1\right)^{1/2}, ~~~~~
\H={1\over 2 \eta}, ~~~~~
\vp' ={\sqrt 3 \over \eta}, ~~~~~
z=+2 \sqrt 3 a, ~~~~~~~
\eta \geq +\eta_1.
\label{619}
\eea
We will concentrate our explicit example on the computation of the
scalar spectrum outside the horizon, and we will start our discussion
by assuming a smooth behaviour of the curvature perturbation $\R$
across the bounce. In other words, the matching from
$-\eta_1$ to $\eta_1$ will be first performed by imposing the
continuity of $\R$ and $\R'$. 

We thus consider curvature perturbations of low enough momentum,
outside the horizon, before the bouncing transition,  and we solve
(the Fourier transformed version of ) Eq. (\ref{67}) for $\eta_{ex} \leq
\eta \leq -\eta_1$, where $|k \eta_{ex}|=1$. By using for $v_k$ the
asymptotic expansion in the regime $|k\eta|\ll 1$ \cite{14} (see also
\cite{1}) we obtain
\bea
&&
\R_k(\eta)= A_k\left[1-k^2\int _{\eta_{ex}}^{\eta}z^{-2}d\eta'
\int_{\eta_{ex}}^{\eta'} z^2d\eta''\right]
+kB_k\int_{\eta_{ex}}^{\eta} z^{-2} d\eta' + \cdots ,
\label{620}\\
&&
z^2\R_k'(\eta)= kB_k\left[1-k^2\int _{\eta_{ex}}^{\eta}z^{2}d\eta'
\int_{\eta_{ex}}^{\eta'} z^{-2}d\eta''\right]
-k^2A_k\int_{\eta_{ex}}^{\eta} z^{2} d\eta' + \cdots ,
\label{621}
\eea
where $A_k,B_k$ are integration constants to be determined by the
initial conditions at $\eta=\eta_{ex}$. Such initial conditions, on the
other hand, are prescribed by the canonical normalization (\ref{68}) of
the variable $v_k$, and the result is 
\beq
A_k= {e^{i\a} \over z_{ex} \sqrt{2k}}, ~~~~~~~~~~~
B_k= {z_{ex} \over  \sqrt{2k}} e^{i\b},
\label{622}
\eeq
where $\a$ and $\b$ are appropriate phases, originating from
(possibly) random initial conditions, and where $z_{ex}=-2\sqrt 3 /
\sqrt{k\eta_1}$ (using the background (\ref{618})). The integration of
Eqs. (\ref{620}), (\ref{621}) gives then, to lowest order, 
\beq
\R_k(\eta)=\sqrt \eta_1 \left[c_1 + c_2 \ln (-k\eta)\right], 
~~~~~ \R_k'(\eta)= c_3 \sqrt \eta_1 (-\eta)^{-1},
\label{623}
\eeq
where $c_1,c_2,c_3$ are $k$-independent complex numbers. It should
be recalled that we are working in units $\Mp^2=2$, and that the
correct canonical dimensions of the Fourier modes, $[\R_k]=k^{-3/2}$,
are easily restored  when  Eq. (\ref{623}) is multiplied by $\sqrt 2/\Mp$. 

In order to obtain the spectrum of low-energy modes after the bounce,
but before they re-enter inside the horizon, we will now solve Eq.
(\ref{67}) for $\eta_1\leq \eta \leq \eta_{re}$, where $k\eta_{re}=1$,
by using the values of $\R$ and $\R'$ computed at $\eta=-\eta_1$, from
eq. (\ref{623}), as initial conditions at $\eta=\eta_1$. The asymptotic
expansions (\ref{620}), (\ref{621}) are still valid, with the only
difference that the lower limit of integration is now $\eta_1$ instead
of $\eta_{ex}$. The integration constants $A_k,B_k$ are now
determined by the conditions
\bea
&&
\R_k(\eta_1)=A_k= \R_k(-\eta_1)= 
\sqrt \eta_1 \left[c_1 + c_2 \ln (k\eta_1)\right], 
\label{624a}\\
&&
\R_k'(\eta_1)={k\over z_1^2} B_k= \R_k'(-\eta_1)= c_3/ \sqrt{\eta_1}.
\label{624}
\eea
Using the background (\ref{619}) we thus get, to lowest order, 
\bea
&&
\R_k(\eta)=\sqrt \eta_1 \left[c_1 + c_2 \ln (k\eta_1)+c_3 \ln
(k_1\eta)\right],  \nonumber\\
&&
z^2\R_k'(\eta)= {1\over \sqrt \eta_1}\left[ 12 c_3 -{1\over 2}
(k\eta)^2 \ln (k\eta_1)\right], ~~~~~
\eta_1\leq \eta \leq \eta_{re}, ~~~~ k\eta \ll 1,\label{625}
\eea
which provides the spectrum of curvature perturbations outside the
horizon, after the bounce, in agreement with our expectations
(\ref{616}). The corresponding spectrum for the Bardeen potential is
then obtained from Eq. (\ref{69}), which leads to
\beq
\Psi_k(\eta)={\H \over 4 k^2 a^2} z^2 \R_k'=
{\sqrt \eta_1 \over (k\eta)^2}\left[{3\over 2} c_3 -{(k\eta)^2\over 16} 
\ln (k\eta_1)\right].
\label{626}
\eeq
The leading term, for $k\eta \ll1$, is in agreement with the
expectation (\ref{617}). 

The above results for the scalar perturbation spectrum, obtained
through a matching procedure based on the continuity of $\R$ and
$\R'$, will be confirmed by the numerical computations of the next
subsection. Let us now show that a matching based on the continuity of
$\Psi$ and $\Psi'$ leads to  different results for the scalar spectrum. 

Following the same procedure as before, we start considering the
low-energy modes of the Bardeen potential outside the horizon, before
the transition, and we solve  Eq. (\ref{63}) for $\eta_{ex}\leq \eta \leq
-\eta_1$. By using the background (\ref{618}) we are lead to the
asymptotic expansion
\bea
&&
\Psi_k(\eta)= -{\eta_1\over 4 \eta^2}\left[A_k\left(1-k^2\int
_{\eta_{ex}}^{\eta}Z^{-2}d\eta' \int_{\eta_{ex}}^{\eta'}
Z^2d\eta''\right)
+kB_k\int_{\eta_{ex}}^{\eta} Z^{-2} d\eta' \right],
\label{627}\\
&&
\Psi_k'(\eta)={\eta_1\over 2\eta^3}A_k\left[1-k^2\int
_{\eta_{ex}}^{\eta}Z^{-2}d\eta' \int_{\eta_{ex}}^{\eta'}
Z^2d\eta'' 
+{k^2\eta\over 2 Z^2} \int_{\eta_{ex}}^{\eta} Z^{2} d\eta' \right] 
\nonumber\\
&&
+{\eta_1\over 2\eta^3}B_k\left[k\int_{\eta_{ex}}^{\eta} Z^{-2} d\eta' 
-{k\eta\over 2 Z^2}\left(1-k^2\int
_{\eta_{ex}}^{\eta}Z^{2}d\eta' \int_{\eta_{ex}}^{\eta'}
Z^{-2}d\eta'' \right)
\right],
\label{628}
\eea
where $A_k, B_k$ are determined by the normalization of $U_k$ at
$\eta=\eta_{ex}$, namely $|U_k|=|\xi_{ex} \Psi_k(\eta_{ex})|=k^{-3/2}$. The
result is
\beq
A_k= {2 \sqrt 3 \over k^2 \sqrt \eta_1} e^{i\a}, ~~~~~~~~~
B_k= {\sqrt \eta_1 \over k \sqrt 3} \left( e^{i\a} -{1\over 2} e^{i
\b}\right).
\label{629}
\eeq
We thus obtain, to lowest order, 
\beq
\Psi_k(\eta)= {b_1\sqrt \eta_1 \over (k\eta)^2}, ~~~~~~
\Psi_k'(\eta)= {b_2\sqrt \eta_1 \over k^2\eta^3}, ~~~~~~~~~
\eta \leq -\eta_1,
\label{630}
\eeq
where $b_1,b_2$ are $k$-independent complex numbers. It may be
noted that, {\em before} the bounce,  the spectral distribution of
$\Psi_k$ is in agreement with that we would obtain from the
spectrum of $\R_k$, eq. (\ref{623}), by using the general relation
(\ref{617}). 

Let us now compute $\Psi_k$ after the bounce, by solving eq.
(\ref{63}) for $\eta_1 \leq \eta \leq \eta_{re}$ before re-entering, and
using $\Psi_k(-\eta_1)$ and $\Psi_k'(-\eta_1)$ as initial conditions at
$\eta=\eta_1$. The asymptotic expansion after the bounce, using the
background (\ref{619}), and taking into account all the required flips of
sign, takes the form:
\bea
&&
\Psi_k(\eta)= {\eta_1\over 4 \eta^2}\left[A_k\left(1-k^2\int
_{\eta_{1}}^{\eta}Z^{-2}d\eta' \int_{\eta_{1}}^{\eta'}
Z^2d\eta''\right) 
+kB_k\int_{\eta_{1}}^{\eta} Z^{-2} d\eta' \right],
\label{631}\\
&&
\Psi_k'(\eta)=-{\eta_1\over 2\eta^3}A_k\left[1-k^2\int
_{\eta_{1}}^{\eta}Z^{-2}d\eta' \int_{\eta_{1}}^{\eta'}
Z^2d\eta'' 
-{k^2\eta\over 2 Z^2} \int_{\eta_{1}}^{\eta} Z^{2} d\eta' \right] 
\nonumber\\
&&
-{\eta_1\over 2\eta^3}B_k\left[k\int_{\eta_{1}}^{\eta} Z^{-2} d\eta' 
+{k\eta\over 2 Z^2}\left(1-k^2\int
_{\eta_{1}}^{\eta}Z^{2}d\eta' \int_{\eta_{1}}^{\eta'}
Z^{-2}d\eta'' \right)
\right].
\label{632}
\eea
The integration constants are determined by the continuity of $\Psi$
and $\Psi'$ as
\beq
A_k= {4 b_1 \eta_1^{3/2} \over (k\eta_1)^2}, ~~~~~~~~
B_k= -{1\over 3} (2 b_1+b_2){\eta_1^{3/2} \over (k\eta_1)^3}.
\label{633}
\eeq

The super-horizon spectrum of the Bardeen potential, after the bounce, to leading order, is then determined by
\beq
\Psi_k(\eta)= {\sqrt\eta_1 \over (k \eta)^2}\left[b_1- \left(b_1+{b_2\over 2}\right)\left(\eta\over \eta_1\right)^2\right] \sim 
{\sqrt\eta_1 \over (k \eta_1)^2}, ~~~~\eta \gg \eta_1, ~~~~ k\eta \ll 1.
\label{634}
\eeq
This shows that the Bardeen modes, in case of a smooth crossing of  the
bouncing region, would tend to stay constant after the bounce (outside the
horizon), instead of decreasing in time like $\eta^{-2}$, as predicted by Eq.
(\ref{626}). 

Let us now compare the above results with a numerical integration of  the
perturbation equations. 

\subsection{Numerical analysis and comparison with analytic estimates}
\label{Sec5.3}

For a numerical computation of the spectrum it is convenient to express 
the evolution of our perturbations $\R$ and  $\Psi$ directly in the S-frame,
where we know the explicit form of the background solution, Eq. (\ref{333}).
In terms of S-frame variables, using the gauge $\da \vp=0$, we have, in
particular (see Subsection \ref{Sec5.2}), 
\bea &&
\R_E= -\psi_E=-\psi_s, \nonumber\\
&&
\Psi_E=\psi_E+\H_EE_E'= \psi_s +\left(\H_s-{\vp'\over 2}\right) E_s'.
\label{635}
\eea
>From the system of scalar perturbation equations (\ref{512})-(\ref{515}) we can then obtain a system of  first order, coupled equations for the variables $\psi_s$ and $E'$, 
whose numerical solutions will allow us to determine $\R_E$ and $\Psi_E$ everywhere. We will find that $\R$, unlike $\Psi$, goes smoothly through the bouncing transition, 
and that the numerical results provide a good fit of the scalar spectra analytically obtained using the continuity  of $\R$ and $\R'$.

Dropping, for simplicity, the subscript referring to S-frame variables, and using the transformations (\ref{529}), (\ref{531}), 
the perturbation equations  (\ref{58})--(\ref{510}) can be written in the S-frame as 
\bea
&&
E''+ (2 \H -\vp')E'+\psi-\phi=0,
\label{637}\\
&&
2\psi' + (2 \H -\vp') \phi=0,
\label{638}\\
&&
2 \nabla^2\psi + (2 \H -\vp') \nabla^2 E' -3 \psi'  (2 \H -\vp') 
-(6 \H^2 -6 \H \vp' +{\vp'}^2)=0.
\label{639}
\eea
By eliminating $\phi$ through Eq. (\ref{638}), and moving to the S-frame cosmic time coordinate $t$, the combination of the above equations leads to the first-order autonomous system of equations:
\begin{eqnarray}
&& \dot{\psi}_{k} = A_{k}(t) \psi_{k} + B_{k}(t) \Ga_{k},
\label{firsteq}\\
&& \dot\Ga_{k} = C_{k}(t) \psi_{k} + D_{k}(t) \Ga_{k},
\label{seconeq}
\end{eqnarray}
for the variables 
 $\psi_k = - {\cal R}_{k}$ and $\Ga_k=a^2 \dot E_k$ (an overdot denotes 
differentiation with respect to $t$). The coefficients appearing 
in Eqs. (\ref{firsteq}) and (\ref{seconeq}) are
\begin{eqnarray}
A_{k}(t) = \frac{2 (\dot{\vp} - 2  H)}{\dot{\vp} }\omega^2, ~~ &&~~ B_{k}(t) =-\biggl(\frac{\dot{\vp} - 2  H}{\dot{\vp}}\biggr)^2 
\omega^2,
\nonumber\\
C_{k}(t) = \frac{4\omega^2}{ \dot{\vp}^2} -1 ,~~ &&~~ 
D_{k}(t) = \dot{\varphi} -  H  - A_{k}(t).
\label{coeff}
\end{eqnarray}
where $ \omega = k/a$ denotes the physical momentum.
Using eq. (\ref{635}) it is also clear that the E-frame Bardeen potential is given by
\begin{equation}
\Psi^E_{k} = \psi_{k} - \frac{1}{2} (\dot{\vp} -2 H)\Ga_{k}.
\label{bard}
\end{equation}  

In order to solve numerically Eqs. (\ref{firsteq}) and (\ref{seconeq}) in the case of the non-singular model introduced in the previous sections, quantum mechanical initial conditions are to be 
imposed at early enough times, when the relevant mode is still inside the horizon. We define 
\bea
&&
{\cal R}_{k} = F_{k} + i Q_{k},~~~~~~~~~~~~
\dot{{\cal R}}_{k} = G_{k} + i P_{k},\\
&&
w(t) = \sqrt{a} z =  - 2 \frac{ \dot{\vp}}{\dot{\vp} -2 H} a^{3/2} e^{-\vp/2},
\eea
and we use the canonical normalization (\ref{68}). In particular, we choose the initial phase at $t=t_i$ in such a way that the initial value of $\R_k$ is real, namely 
\begin{eqnarray}
&& F_{k}(t_{ i}) = \frac{1}{w_i \sqrt{2 \omega_i} },
~~~~~~~~~~~~~~~~~~~~~~~~~ Q_{k}(t_{ i})= 0,
\nonumber\\
&& G_{k} (t_{i})= \biggl(\frac{H}{2} - \frac{\dot{w}}{w}\biggr)_{t_i}
 ~\frac{1}{w_i \sqrt{2 \omega_i}},~~~~~~~~
  P_{k}(t_{ i}) = - \sqrt{\frac{\omega_i}{2}} \frac{1}{w_i}.
\label{QM}
\end{eqnarray}
Using (\ref{firsteq}), (\ref{seconeq})  the initial conditions for the real 
and imaginary parts of $\Ga_{k}$ are then fixed as follows:
\begin{eqnarray}
&& 
{\rm Re}\{\Ga_{k}(t_{ i})\} = -\left(G_k-A_kF_k\over B_k\right)_{t_i}, 
\nonumber\\
&& 
{\rm Im}\{\Ga_{k}(t_{ i})\} = -\left(P_k-A_kQ_k\over B_k\right)_{t_i}. 
\end{eqnarray}

The spectrum of all the interesting quantities can now be obtained 
through a numerical integration of Eqs. (\ref{firsteq}), (\ref{seconeq}). 
Consider, for instance, 
the spectrum of curvature perturbations  $\delta_{\cal R}= k^{3/2} |{\cal R}_{k}|$. In Fig. \ref{R1} we have illustrated our numerical results for the super-horizon evolution of $\da_\R$  for different comoving momenta,
normalized to a quantum spectrum of vacuum fluctuations. 
We have used the background solution (\ref{333}) with $t_0=1$, and we have fixed the absolute amplitude of the normalized spectrum by choosing $\vp_0$ in such a way that
\beq
\left(H\over \Mp\right)_{t=0} = {e^{\vp_0/2}\over \sqrt 3 ~t_0}=10^{-2}.
\label{norma}
\eeq
The mode $\kappa \equiv k t_0 \sim 1$ starts its first oscillation already for
$t \sim 0$, so that it basically exits the horizon and then immediately 
re-enters. 
Smaller wave-numbers stay longer outside the horizon. We have only considered low-fequency modes, $\kappa \ll1$, which are consistently amplified in the given model of bouncing background. Higher frequency modes may develop a quantum gravitational instability (near the origin $t=0$, where $c_s^2<0$), similar to that of tensor modes in higher-derivative string cosmology backgrounds \cite{15}. For such modes our (low-energy) model of bakground is clearly inadequate. 

\begin{figure}
\centerline{\epsfxsize = 10cm  \epsffile{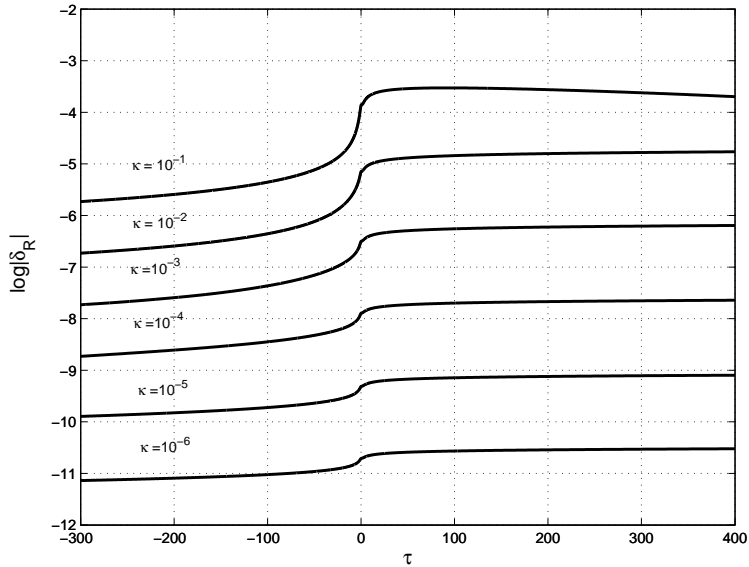}}
\vskip 3mm
\caption[a]{Numerical results for the time-evolution of $\da_\R$, and 
for different values of the comoving wave-number (we have set $t_0=1$, and used the normalization (\ref{norma})).}
\label{R1} 
\end{figure}

>From Fig. \ref{R1} it is already apparent that the spectrum of $\cal{R}$, outside the horizon, follows the predicted power-law distribution $k^{3/2}$, in agreement with Eqs. (\ref{616}) and (\ref{625}). 
For a more accurate determination of the spectrum we will choose a fiducial time $\tau_x=100$, for which all modes with $\kappa \leq 10^{-2}$ are still outside the horizon. The spectrum $|\da_\R(\tau_x)|$ can then be 
computed as a function of $\kappa$,  and the obtained result 
can be fitted with the analytical expression of $\delta {\cal R}_{k}$ obtained 
with the  matching procedure outlined in the previous subsection. 

For numerical reasons, it is convenient to work with the squared modulus of Eq. (\ref{625}) (since the coefficients appearing in that equation are complex). We thus set, at $\tau=\tau_x$, 
\begin{equation}
|\delta_{\cal R}(\tau_x)|^2 = \kappa^3  ( \alpha_{1} + \beta_{1}\ln{\kappa} + \gamma_{1}\ln^2{\kappa}),
\label{dr}
\end{equation} 
where $\alpha_1$, $\beta_1$ and $\gamma_1$ are real numbers.
In Fig. \ref{R2} the  numerical results for the values of $\delta_\R$ are reported with the open diamonds,
 for a selection of modes $10^{-2} \leq \kappa \leq  10^{-6}$, and $\tau_x=100$. Such values are well fitted by Eq. (\ref{dr}) with
\begin{equation}
\alpha_1 = 0.339\times 10^{-4},~~~~~~\beta_1 = - 0.345 \times 10^{-4} , ~~~~~~\gamma_{1} = 0.664\times 10^{-6}.
\label{param}
\end{equation}
The corresponding curve (plotted in Fig. \ref{R2} with a dashed line)
nicely intepolates through the numerical points, confirming the previous results for the analytical estimate of  the spectrum of ${\cal R}_{k}$. 
Note that, when the spectral amplitude is expressed in terms of the correct dimensional factors, we obtain (in our model of matching) $\a_1 \sim \b_1 \sim (H_1/\Mp)^2$, where $H_1 = H(\eta_1)$. Eq. (\ref{param}) thus represents a fit of the matching scale $H_1$, in agreement with the assumed normalization (\ref{norma}). 

\begin{figure}
\centerline{\epsfxsize = 10cm  \epsffile{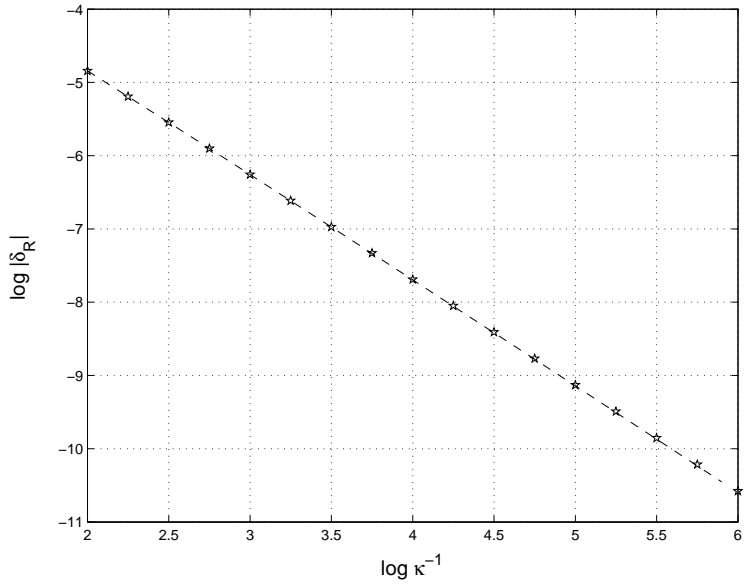}}
\vskip 3mm
\caption[a]{Spectral distribution of $|\da_\R|$, at the fixed 
fiducial time $\tau_{ x}=100$ when all the selection of modes is well outside the horizon. The dashed line 
is the fit given in Eqs. (\ref{dr}) and (\ref{param}) (we are using decimal logarithms on both axes).}
\label{R2} 
\end{figure}

A similar numerical analysis can now be repeated for the spectrum of the E-frame Bardeen potential, $|\da_\Psi|$, from Eq. (\ref{bard}). The numerical solutions show, first of all, that the values of $\Psi$ and $\Psi'$ (though continuous at $t=0$) change drastically while crossing the bouncing region (unlike $\R$), as illustrated in Fig.  \ref{PSI1}. This support the expectation that the correct spectral distribution should be obtained, analytically, by matching the curvature perturbation variable, instead of the Bardeen potential. 

\begin{figure}
\centerline{\epsfxsize = 10cm  \epsffile{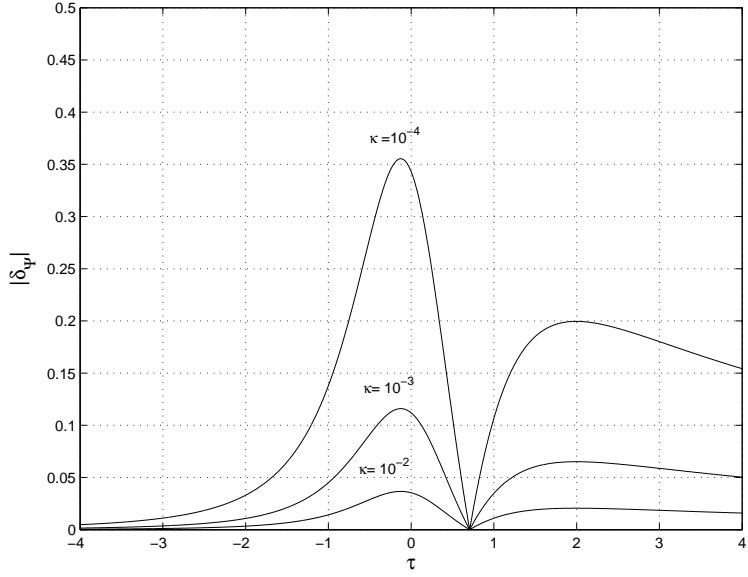}}
\vskip 3mm
\caption[a]{Evolution of the (E-frame) Bardeen spectrum around the bounce.}
\label{PSI1} 
\end{figure}

This expectation is indeed confirmed by Fig. \ref{PSI2}, where we plot the Bardeen spectrum, as a function of time, for different comoving momenta which are all inside the horizon. The ``post-bounce" evolution of the given super-horizon modes, instead of being time-independent  as required by Eq. (\ref{634}), decreases in time exactly as predicted by Eq. (\ref{626}), namely $\Psi_k(\eta) \sim \eta^{-2}$. Sufficiently far from the bounce, where the S-frame cosmic time $t$ is related to the conformal time $\eta$ by
\beq
t= \int a_s d\eta= \int a_E e^{\vp/2} d \eta \sim \eta^{(3 + \sqrt 3)/ 2}
\label{cosmic}
\eeq
(we have used eq. (\ref{612})), we thus obtain 
\beq
\Psi_k \sim \eta^{-2} \sim t^{-4/(3 + \sqrt 3)},  ~~~~~~~~
4/(3 + \sqrt 3)\simeq 0.845.
\eeq
This general behaviour predicted by Eq. (\ref{626}) is represented in Fig. 
\ref{PSI2} by the dashed curve, which nicely agrees with the time-evolution of the Bardeen spectrum for all plotted (super-horizon) modes. 

\begin{figure}
\centerline{\epsfxsize = 10cm  \epsffile{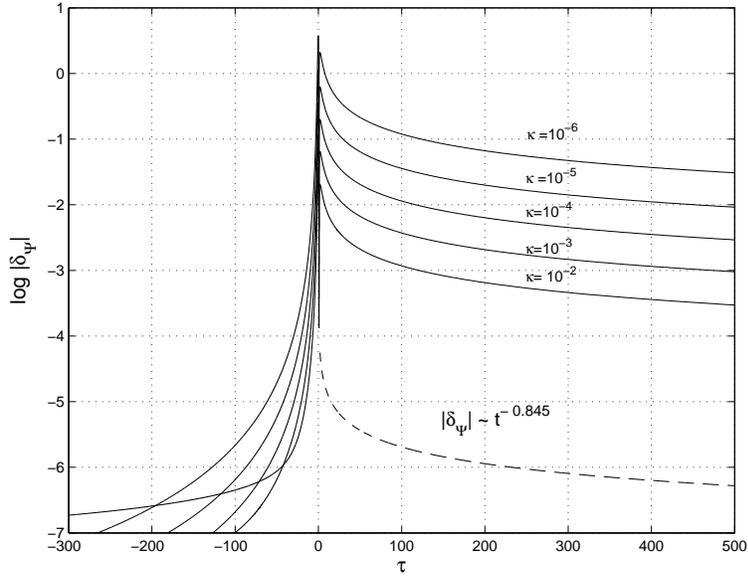}}
\vskip 3mm
\caption[a]{Evolution of the (E-frame) Bardeen spectrum for super-horizon modes, after the bounce. The dashed curve parametrizes (with an arbitrary normalization) the time behaviour $\eta^{-2} \sim t^{-4/(3 + \sqrt 3)}\simeq t^{-0.845}$ predicted by Eq. (\ref{626}).}
\label{PSI2} 
\end{figure}

Also in agreement with the analytical prediction (\ref{626}) is the spectral distribution of the super-horizon modes reported in Fig. \ref{PSI2} , which can be easily inferred from the (decimal) logarithmic scale of the figure as 
$|\da_\Psi|=k^{3/2} |\psi_k| \sim k^{-1/2}$. For a more precise computation of the spectrum we may observe that the numerical values of $|\da_\Psi|$, at a given fiducial time $\tau_x=100$ such that all the considered modes are still outside the horizon, can be fitted by 
\begin{equation}
\delta \Psi_{k} = \frac{\alpha_{2}}{\kappa^{1/2}}, ~~~~~~~~~~~
\alpha_2 = 0.347 \times 10^{-8}.
\label{fitps}
\end{equation}
Such a numerical fit,  illustrated in Fig. \ref{PSI3}, precisely confirms the leading order distribution analytically predicted in Eq. (\ref{626}). 

\begin{figure}
\centerline{\epsfxsize = 10cm  \epsffile{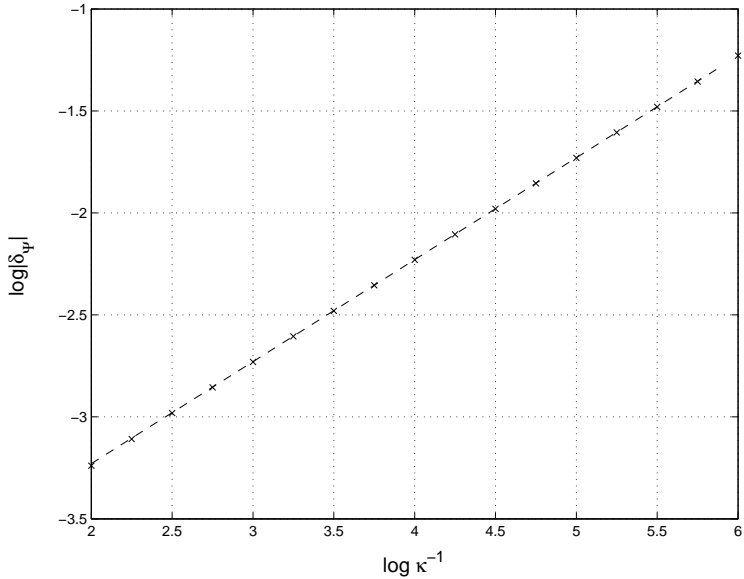}}
\vskip 3mm
\caption[a]{Spectral distribution of $|\da_\Psi|$ at the fixed fiducial 
time $\tau_{\rm x}=100$, when all the modes included in the plot are well outside the horizon. The dashed line is the fit reported in Eq. (\ref{fitps}).}
\label{PSI3} 
\end{figure}

\section{Conclusion}
\label{Sec6}
\setcounter{equation}{0}

If taken at face value  the results presented in this paper  indicate that the curvature perturbation
$ \cal{R}$, rather than the Bardeen potential $\Psi$, behaves smoothly through the bounce.
The same conclusion can be reached by looking at  differential equations
obeyed by $\Psi$ and $ \cal{R}$, Eqs. (\ref{61}) and (\ref{65}), and at their formal solutions Eqs. (\ref{620}) and (\ref{627}), respectively. 
The solutions for $ \cal{R}$ and its canonically conjugate momentum $\Pi_{\cal{R}} \equiv z^2  \cal{R'}$ only contain the
backgrounds (through  $z$) under integration, thus smoothing out their rapid variation at the bounce. By contrast, the solution for $\Psi$ and its time derivative contain, besides integrals of $Z^2$ and $Z^{-2}$,
additional rapidly varying pre-factors. The only way to make $\Psi$ smooth across the bound appears to be
through the insertion of $\delta$-function like contributions to  $Z$ so that the variations of the prefactors
are carefully cancelled by the integrals.  Such  $\delta$-function contributions will instead make $ \cal{R}$ behave wildly through the
bounce.

In particular, the fact that $\Pi_{\cal{R}}$ and $\Psi$ cannot be simultaneously smooth across the bounce stems from the simple relation between them
\beq
k^2 \Psi \sim \H a^{-2} \Pi_{\cal{R}} \; ,
\eeq
(see Eq. (\ref{69})), 
and the fact that the quantity $\H a^{-2}$ clearly changes sign across the bounce (recall that we are using Einstein-frame quantities).
If, on the basis of our example and of the above discussion, one concludes that $\cal{R}$ is continuous across the
bounce, the spectra of adiabatic scalar perturbations in bouncing cosmologies is expected to be far from scale-invariant.
In this case the only known way to rescue the phenomenology of PBB cosmologies is through use of the so-called
``curvaton" mechanism \cite{curvaton} (see for instance \cite{1,21}).

On the other hand, one has to concede the fact that our mechanism for producing the bounce through a non-local potential
tends to tie its occurrence to a critical value of the scalar field, and thus to privilege continuity of perturbations on comoving (constant $\phi$) hypersurfaces. The proponents of a scale-invariant spectrum  \cite{EKPsp} claim that the situation is totally different in the ekpyrotic/cyclic  scenario, provided this is looked at from a  five-dimensional perspective. Possibly, such an approach
will fix the matching conditions in a completely different way:  what appears to be fine-tuning in four-dimensions  could appear to be natural from a higher dimensional point of view.  As a result,  even the spectrum of tensor perturbations
claimed in \cite{9} differs from the one proposed earlier in 
\cite{BGGMV,BGGV} by the lack of a logarithmic 
behaviour of the Bogolubov coefficient.

In conclusion, we believe it would be very interesting to complement the work of  \cite{DV} by linking in a precise way  different matching prescriptions to different dynamical mechanisms for inducing the bounce and avoid the singularity. This will eventually connect physical observables, such as the tilt in the power spectra, to the very non-trivial, and yet poorly understood, dynamics that supposedly allows one to evade the Hawking-Penrose singularity theorems and thus to conceive the possibility of a pre-big bang  Universe.

\vskip 2 cm

\section*{Acknowledgements}

This research was supported in part by the National Science Foundation under Grant No. PHY99-07949. 
Two of us (M. Gasperini and G. Veneziano) are glad to aknowledge the financial support of the {\em ``Superstring Cosmology "} research  program and the kind ospitality of the {\sl Kavli Institute for Theoretical Physics} (University of California, Santa Barbara), where part of this work was carried out.

\end{document}